\documentclass[a4paper,11pt]{article}
\pdfoutput=1 

\usepackage{jcappub} 

\usepackage[T1]{fontenc} 
\usepackage{subfig}
\usepackage[utf8]{inputenc}
\usepackage[english]{babel}
\usepackage{ulem}

\usepackage{xcolor}

\title{\boldmath Intensity and anisotropies of the stochastic Gravitational Wave background from merging compact binaries in galaxies}

\author[a,b,c]{Giulia Capurri,}
\author[a,b,c,d]{Andrea Lapi,}
\author[a,b,c]{Carlo Baccigalupi,}
\author[a,b,c]{Lumen Boco,}
\author[a,b,c]{Giulio Scelfo,}
\author[a,b,c]{Tommaso Ronconi}

\affiliation[a]{SISSA, via Bonomea 265, 34136, Trieste, Italy}
\affiliation[b]{INFN, Sezione di Trieste, via Bonomea 265, 34136, Trieste, Italy}
\affiliation[c]{IFPU, via Beirut 2, 34151, Trieste, Italy}
\affiliation[d]{INAF/OATS, via Tiepolo 11, I-34143 Trieste, Italy}

\emailAdd{giulia.capurri@sissa.it}
\emailAdd{andrea.lapi@sissa.it}
\emailAdd{carlo.baccigalupi@sissa.it}
\emailAdd{lumen.boco@sissa.it}
\emailAdd{giulio.scelfo@sissa.it}
\emailAdd{tommaso.ronconi@sissa.it}

\abstract{We investigate the isotropic and anisotropic components of the Stochastic Gravitational Wave Background (SGWB) originated from unresolved merging compact binaries in galaxies. We base our analysis on an empirical approach to galactic astrophysics that allows to follow the evolution of individual systems. We then characterize the energy density of the SGWB as a tracer of the total matter density, in order to compute the angular power spectrum of anisotropies with the Cosmic Linear Anisotropy Solving System (\texttt{CLASS}) public code in full generality. We obtain predictions for the isotropic energy density and for the angular power spectrum of the SGWB anisotropies, and study the prospect for their observations with advanced Laser Interferometer Gravitational-Wave and Virgo Observatories and with the Einstein Telescope. We identify the contributions coming from different type of sources (binary black holes, binary neutron stars and black hole-neutron star) and from different redshifts. We examine in detail the spectral shape of the energy density for all types of sources, comparing the results for the two detectors. We find that the power spectrum of the SGWB anisotropies behaves like a power law on large angular scales and drops at small scales: we explain this behavior in terms of the redshift distribution of sources that contribute most to the signal, and of the sensitivities of the two detectors. Finally, we simulate a high resolution full sky map of the SGWB starting from the power spectra obtained with \texttt{CLASS} and including Poisson statistics and clustering properties.}

\newcommand{\mc}{\mathcal{M}_{c}}
\newcommand{\omegagw}{\Omega_{\rm{gw}}}
\newcommand{\baromega}{\bar{\Omega}_{\rm{gw}}}
\newcommand{\fobs}{f_{\rm{o}}}
\newcommand{\eobs}{\hat{e}_{\rm{o}}}

\begin{document}
\maketitle
\flushbottom

\section{Introduction}
\label{sec:intro}
The discovery of Gravitational Waves (GWs) by the advanced Laser Interferometer Gravitational - Wave and Virgo Observatories (aLIGO/Virgo) collaboration \citep{Abbott:2016blz,TheLIGOScientific:2016pea,Abbott:2016nmj,Abbott:2017vtc,Abbott:2017oio,TheLIGOScientific:2017qsa,Abbott:2017gyy,LIGOScientific:2018mvr} has opened a new observational window on the Universe. We are currently in the third observing run (O3) and a large number of new GW events from binary black hole (BH-BH), neutron star (NS-NS), and black hole-neutron star (BH-NS) mergers is being detected \citep{Abbott:2020uma,LIGOScientific:2020stg,Abbott:2020khf,Abbott:2020tfl,Abbott:2020niy,Abbott:2020gyp,LIGOScientific:2021qlt}. Together with the observation of single events, the detection and characterization of the Stochastic Gravitational Wave Background (SGWB) from unresolved compact binary coalescences is one of the most challenging targets for the GW community. The search for the isotropic SGWB using the O3 data, combined with upper limits from the earlier O1 and O2 runs \citep{TheLIGOScientific:2016dpb,LIGOScientific:2019vic}, constrains its energy density to be $\omegagw < 5.8 \times 10^{-9}$ at 25 Hz with $95\%$ confidence for a flat SGWB and $\omegagw < 3.4 \times 10^{-9}$ for a power law SGWB with a spectral index of 2/3, consistently with expectations for compact binary coalescences \citep{Abbott:2021upperlimits}. Likewise, within current uncertainties, the data are consistent with an isotropic SWGB \citep{TheLIGOScientific:2016xzw,LIGOScientific:2019gaw, Abbott:2021jel}. 

The phenomena that can produce a SGWB can be classified with respect to their astrophysical or cosmological origin (see \citep{Christensen:2018iqi} for a general review). The astrophysical SGWB is constituted by the incoherent superposition of signals from of a large number of unresolved astrophysical sources, including merging double compact objects, merging supermassive black holes binaries \citep{Jaffe:2002rt,Kelley:2017lek}, core collapse supernovae  \citep{Buonanno:2004tp,Crocker:2015taa,Crocker:2017agi}, rotating neutron stars \citep{Rosado:2012bk,Lasky:2013jfa,Talukder:2014eba} and population III binaries \citep{Inayoshi:2016hco}. On the other hand, cosmological contributions to the SGWB are primarily expected from the amplification of vacuum quantum fluctuations during inflation \citep{Barnaby:2011qe,Cook:2011hg,Guzzetti:2016mkm}, but also from GWs produced during the reheating phase at the end of inflation itself \citep{Easther:2006gt,Easther:2006vd,GarciaBellido:2007dg}. Inflationary scenarios, also involving recently proposed models based on SU(2) symmetries, can be constrained from GWs observatories in conjunction with B-modes of the Cosmic Microwave Background observations \citep{Campeti:2020xwn}. Other, perhaps more exotic, cosmological sources of GWs are pre-Big Bang scenarios \citep{Gasperini:1992em,Mandic:2005bd,Gasperini:2016gre}, cosmic strings \citep{Damour:2004kw,Siemens:2006yp,Seljak:2006hi}, primordial black holes \citep{Sasaki:2018dmp,Nakama:2016gzw,Bartolo:2019zvb}, first-order phase transitions in the early universe \citep{PhysRevLett.69.2026,Kamionkowski:1993fg,Baccigalupi:1997re,Binetruy:2012ze} and magnetic fields \citep{Caprini:2001nb,Paoletti:2019pdi}. The anisotropies of the cosmological SGWB have been characterized in many recent works (see e.g. \citep{Bartolo:2019oiq,Bartolo:2019yeu,DallArmi:2020dar}). Different GW backgrounds contribute at different frequencies and have distinct statistical properties, so that their characterization potentially allows to distinguish them from possible future detections.

The SGWB will be scrutinized by powerful observatories in the forthcoming years and decades. 
The current generation of ground-based laser interferometers (LIGO \citep{PhysRevD.102.062003}, Virgo \citep{TheVirgo:2014hva}, the Kamioka Gravitational Wave Detector, KAGRA, \citep{Somiya:2011np}) will be succeeded by Einstein Telescope (ET) \citep{Sathyaprakash:2012jk,Regimbau:2012ir,Maggiore:2019uih,Hild:2010id,Mentasti:2020yyd} and Cosmic Explorer (CE) \citep{Reitze:2019iox}, both operating between a few Hertz and a few kilo-Hertz.
The Laser Interferometer Space Antenna (LISA) is scheduled for observations in the early 2030s, targeting a variety of SGWB sources in the milli-Hertz band (see e.g. \citep{Bartolo:2016ami,Babak:2017tow,Bartolo:2018qqn} and references therein). Other proposals for future space missions include $\mu$Ares \citep{Sesana:2019vho} in the micro-Hertz band; the Advanced Millihertz Gravitational-wave Observatory (AMIGO) \citep{Baibhav:2019rsa} in the milli-Hertz band; the Big Bang Observer (BBO) \citep{Crowder:2005nr,Smith:2016jqs}, the Deci-hertz Interferometer Gravitational wave Observatory (DECIGO) \citep{Seto:2001qf,Kawamura:2020pcg}, the Decihertz Observatory (DO) \citep{Sedda:2019uro} and the Atomic Experiment for Dark Matter and Gravity Exploration in Space (AEDGE) \citep{Bertoldi:2019tck} in the deci-Hertz band. Finally, current generation of pulsar timing arrays (PTA) experiments, such as the Nanohertz Observatory for Gravitational Waves (NANOGrav) \citep{Arzoumanian:2020vkk}, the European PTA \citep{Lentati:2015qwp} and the Parkes PTA \citep{Lasky:2015lej} are placing limits on the SGWB in the nano-Hertz band. In the next future, also Square Kilometre Array (SKA) \citep{Bull:2018lat} will join this international network of PTA.

The SGWB produced by merging Double Compact Objects (DCOs) in galaxies is expected to be one of the dominant contributions to the total background. 
Its characterization is complex since it requires to model many astrophysical and cosmological processes occurring on different times and spatial scales: from stellar physics, through galaxy formation and evolution, up to GW physics and evolution of cosmological perturbations. 
On the other hand, the information richness encoded in the SGWB produced by merging DCOs makes its detection and characterization one of the main goals of GW experiments. For this reason, knowing how this background is related to astrophysical models and parameters will be of paramount importance for improving our understanding of such astrophysical processes once the SGWB will be detected. Moreover, a high precision modelling of the SGWB from astrophysical sources is needed in order to measure other components of the background, as the primordial one expected from cosmic inflation.    

The isotropic component of the SGWB produced by merging DCOs has been computed in many works (e.g. \citep{Regimbau:2011rp,Rosado:2011kv,Marassi:2011si,Zhu:2011bd,Zhu:2012xw,Wu:2011ac,KowalskaLeszczynska2015,TheLIGOScientific:2016wyq,Abbott:2017xzg, Perigois:2020ymr}), according to different prescriptions for stellar and galactic physics. In particular, these studies typically combine population synthesis either with cosmological simulations or with recipes on the cosmic star formation rate density and metallicity distributions inferred from observations.
For what concerns the anisotropies of the astrophysical SGWB, their first theoretical expression was derived in \citep{Cusin:2017fwz,Cusin:2017mjm}. Predictions for the angular power spectrum of the anisotropies based on this expression have been presented in \citep{Cusin:2018rsq,Cusin:2019jpv,Cusin:2019jhg,Alonso:2020mva}, where the influence of various astrophysical models on the power spectrum was also investigated.
Based on the same framework, at least two other approaches have been proposed to compute the anisotropies of the astrophysical SGWB. One of them is treated in \citep{Jenkins:2018nty,Jenkins:2018uac} and the related results for the angular power spectrum are discussed in \citep{Jenkins:2018kxc, Jenkins:2019nks,Jenkins:2019uzp}. A third approach was presented in \citep{Bertacca:2019fnt}, where a new derivation of the SGWB anisotropies in a cosmological context is proposed: in this work, the observed angular power spectrum is computed accounting for all the projection effects intervening between the source and the observer. All the aforementioned results concerning the SGWB anisotropies are compared in a recent work \citep{Pitrou:2019rjz}. There, the authors show that the predictions are basically all equivalent and that they all contain the same well-known cosmological effects, so that differences in predictions can only arise in the way galactic and sub-galactic physics is computed.

In the present paper, we we focus on the isotropic and the anisotropic components of the SGWB produced by merging DCOs in galaxies as possibly detected by aLIGO/Virgo and in the future by the ET. Our analysis is built on an empirical approach to galactic astrophysics, which allows to follow the evolution of individual systems \citep{Boco:2019teq, Scelfo:2020jyw, Boco:2020pgp}. We compute the power spectrum of the SGWB anisotropies in full generality with the public code \texttt{CLASS}, exploiting the formalism similarities between galaxy number counts and SGWB \citep{lesgourgues2011cosmic, Blas_2011}. In order to do so, we characterize the energy density of the SGWB as a tracer of matter density. Finally, we develop a methodology to simulate a stochastic realization of the full-sky map of the SGWB, taking into account the intrinsic Poisson nature of the signal as well as its clustering properties.

The paper is structured as follows: in Section \ref{sec:astro} we describe the physical prescriptions employed to describe the astrophysical sources; in Section \ref{sec:asgwb} we summarize the theory behind the study of the SGWB, from the computation of the isotropic energy density to the formalism behind the calculation of the angular power spectrum of its anisotropies; in Section \ref{sec:results} we characterize the SGWB energy density as a tracer of the total matter density by deriving expressions for its redshift distribution, bias and magnification bias, we present our results for its amplitude and anisotropies and we produce a simulated full-sky map of the expected signal; in Section \ref{sec:conclusions} we draw our conclusions and outline future prospects. 
Throughout this work we adopt the standard flat $\Lambda$CDM cosmology with parameter values from the Planck 2018 legacy release \cite{Aghanim:2018eyx}, with Hubble rate today corresponding to $H_{0}= 67.4$ km s$^{-1}$ Mpc$^{-1}$, Cold Dark Matter (CDM) and baryon abundances with respect to the critical density corresponding to $\Omega_{\rm{CDM}}h^{2}= 0.120$ and $\Omega_{b}h^{2}=  0.022$, respectively, reionization optical depth $\tau=0.054$, amplitude and spectral index of primordial scalar perturbations corresponding to ln$(10^{10}A_{S})=3.045$ and $n_{S}=0.965$, respectively.

\section{Astrophysical sources} 
\label{sec:astro}
In this work we focus on the SGWB produced by the superposition of GW signals coming from a particular class of astrophysical sources: Double Compact Object (DCO) merging binaries. The large amount of astrophysical and cosmological information that it encodes, together with the fact that it is expected to be one of the the dominant contributions to the total SGWB, makes the modelling of this component of the astrophysical SGWB of great scientific interest. In the characterization of the population of DCOs and in the computation of their merger rates, we follow the approach presented in \citep{Boco:2020pgp,Boco:2019teq,Scelfo:2020jyw}, which is briefly sketched hereafter. The merger rate per unit volume $V$ and chirp mass $\mathcal{M}_c$ can be computed as \citep{Barrett:2017fcw,Neijssel_2019}:
\begin{equation} \label{eq:merger_rate}
    \dfrac{d^{2} \dot{N}}{dV d\mc} (t) = \int dt_{d} \int dZ \dfrac{d^{3}N}{dM_{\rm{SFR}} \, d\mc \, dt_{d}} (\mc, t_{d}|Z) \; \dfrac{d^{2}\dot{M}_{\rm{SFR}}}{dVdZ} (Z,t-t_{d})\,,
\end{equation}
where $t$ is the cosmic time, equivalent to redshift, $M_{\rm{SFR}}$ is the star formed mass, $t_{d}$ is the delay time between the formation of the progenitor binary and the merging of the compact objects binary, $Z$ is the galaxy metallicity and $V$ the comoving volume. 
The first term in the integral is related to stellar and binary evolution and represents the number of DCO mergers per unit of star forming mass per bin of chirp mass and time delay. It can be decomposed into three factors:
\begin{equation} \label{eq:star_term}
    \dfrac{d^{3}N}{dM_{\rm{SFR}} \, d\mc \, dt_{d}} (Z) = \dfrac{dN}{dM_{\rm{SFR}}}(Z) \; \dfrac{dp}{d\mc}(Z) \; \dfrac{dp}{dt_{d}}\,,
\end{equation}
where $dN/dM_{\rm{SFR}}$ is the number of merging DCOs per unit of mass formed in stars at metallicity $Z$, $dp/d\mc$ is the metallicity-dependent chirp mass distribution and $dp/dt_{d}$ is the normalized distribution of delay times between the formation of the progenitor binary and the merger. For the former two factors, we rely on the results of the STARTRACK population synthesis simulations\footnote{We use the simulation data publicly available under this url:https://www.syntheticuniverse.org/}, `reference B' model in \citep{Chruslinska:2017odi,Chruslinska:2018hrb}. 
In principle, the STARTRACK code would provide also a delay time distribution, which might be a function of both chirp mass and metallicity. However, a number of studies based on simulations \citep{Chruslinska:2017odi,dominik+12, giacobbo+18, Chruslinska:2018hrb} and observations \citep{Maoz:2013hna} suggest that the dependence on these two parameters is weak and that the distribution is proportional to $t_{d}^{-1}$, normalized to unity between a minimum value $t_{d,\rm{min}} \sim 10^{7} - 10^{8}$ yr and the age of the universe (see e.g. \citep{dominik+12,Chruslinska:2018hrb} for STARTRACK). The value of $t_{d,\rm{min}}$ and the subsequent normalization of the distribution is highly uncertain since it strongly depends on the model prescriptions. For these reasons, we assume $dp/dt_{d}\propto t_{d}^{-1}$ with $t_{d,\rm{min}} = 50$ Myr, independently on metallicity and DCO type. Still, we caveat that this is an approximation, even if corroborated by the aforementioned works, and that in principle we should use the distribution retrieved from the code.  

The second term in the integral of equation \eqref{eq:merger_rate} is related to galaxy evolution and represents the star forming mass per unit time, comoving volume and metallicity. In the present work, we compute it using observationally derived prescriptions. In particular, following the procedure depicted in \citep{Boco:2020pgp}, we exploit the Star Formation Rate Functions (SFRF) as galaxy statistics and the Fundamental Metallicity Relation (FMR) to assign metallicity to galaxies. The SFRF $d^{2}N/d\log\psi/dV$ represents the galaxy number density per logarithmic bin of SFR at different redshifts and have been derived in \citep{Mancuso:2016zrd} from the galaxy Ultra-Violet (UV), Infra-Red (IR), submm, and radio luminosity functions at different redshift. The gas phase metallicity $Z$ of the interstellar medium, instead, is derived through the empirical FMR, see e.g. \citep{Mannucci_2010,Mannucci_2011,Hunt_2016,Curti_2019}, which is a three parameter relation among the stellar mass of the galaxy $M_{\star}$, SFR, and $Z$. Since galaxies are parametrized and counted through the SFRF statistics, we need to relate the SFR to $M_\star$ in order to derive a metallicity. We do it through the main sequence of star forming galaxies, an observational redshift dependent power law relationship between the stellar mass and the SFR (see e.g. \cite{daddi+07, rodighiero+11, rodighiero+15, whitaker+14, speagle+14, schreiber+15, dunlop+17}). Even though the main sequence is a relation followed by most of the star forming objects, there is a non negligible population of galaxies, called starbursts, which lies above the main sequence: their SFR is higher with respect to the SFR predicted by the main sequence at a given mass. In order to model this population, we follow \cite{Boco:2020pgp} that provides a $M_\star$ distribution $dp/dM_{\star} (M_{\star}|z,\psi)$ for galaxies at fixed value of SFR.   

With all these ingredients, we can compute the galactic term of equation \eqref{eq:merger_rate} as:
\begin{equation}
    \dfrac{d^{2}\dot{M}_{\rm{SFR}}}{dVdZ} (Z|z) = \int d \log \psi \, \psi \dfrac{d^{2}N}{dV d\log \psi} \int dM_{\star} \dfrac{dp}{dM_{\star}} (M_{\star}|z,\psi) \dfrac{dp}{dZ} \biggl|_{\rm{FMR}} (Z|Z_{\rm{FMR}}(M_{\star},\psi))\,,
\end{equation}
where $dp/ d\log Z|_{\rm{FRM}} \propto$ $\exp [-(\log Z - \log Z_{\rm{FMR}}(M_{\star},\psi))^{2}/2\sigma^{2}_{\rm{FMR}}]$ is a log normal distribution around the logarithmic metallicity value set by the FMR at fixed stellar mass and SFR. 

\begin{figure}
\centering
\subfloat[]
{\includegraphics[width=.49\textwidth]{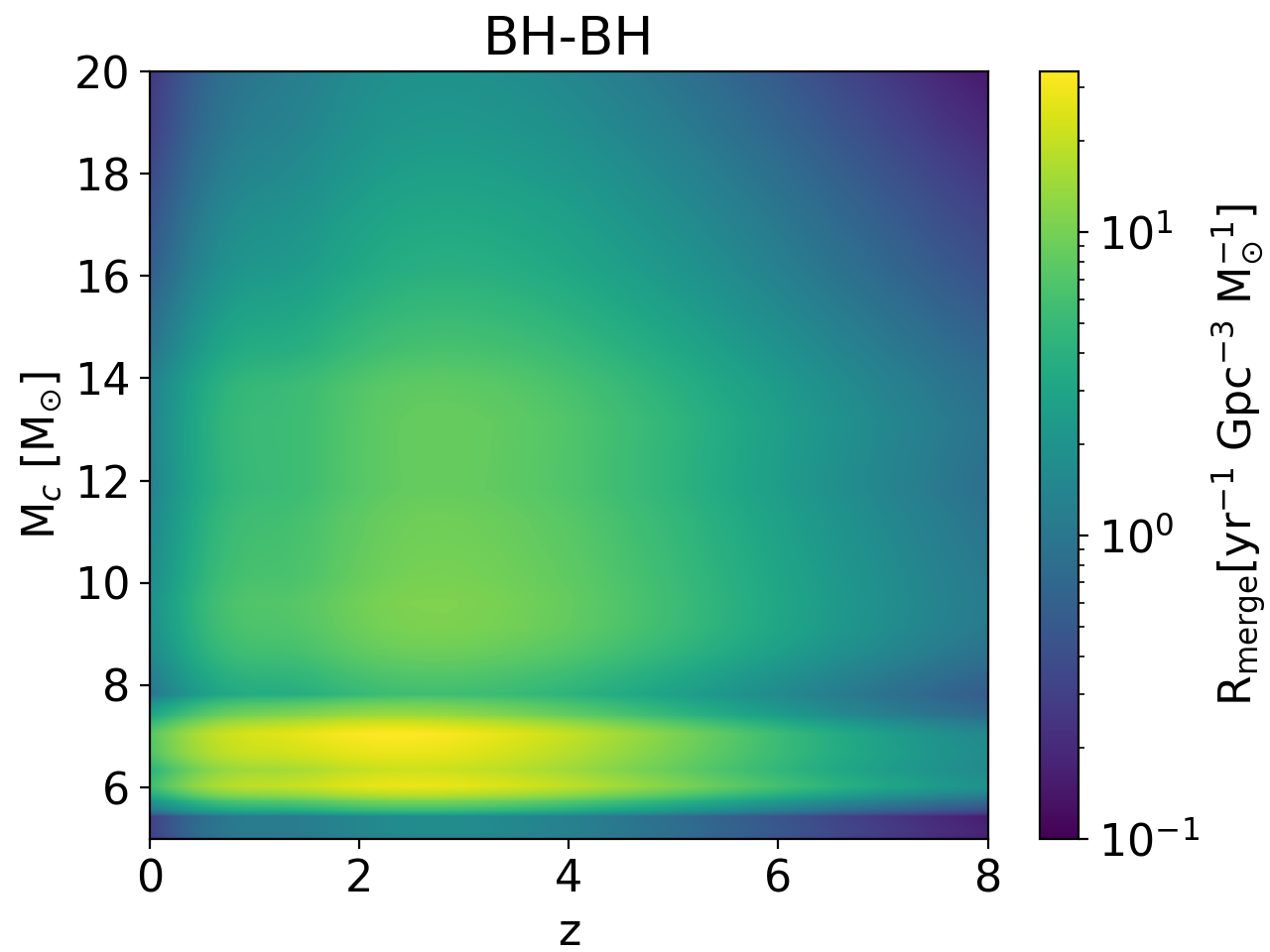}} \ 
\subfloat[]
{\includegraphics[width=.49\textwidth]{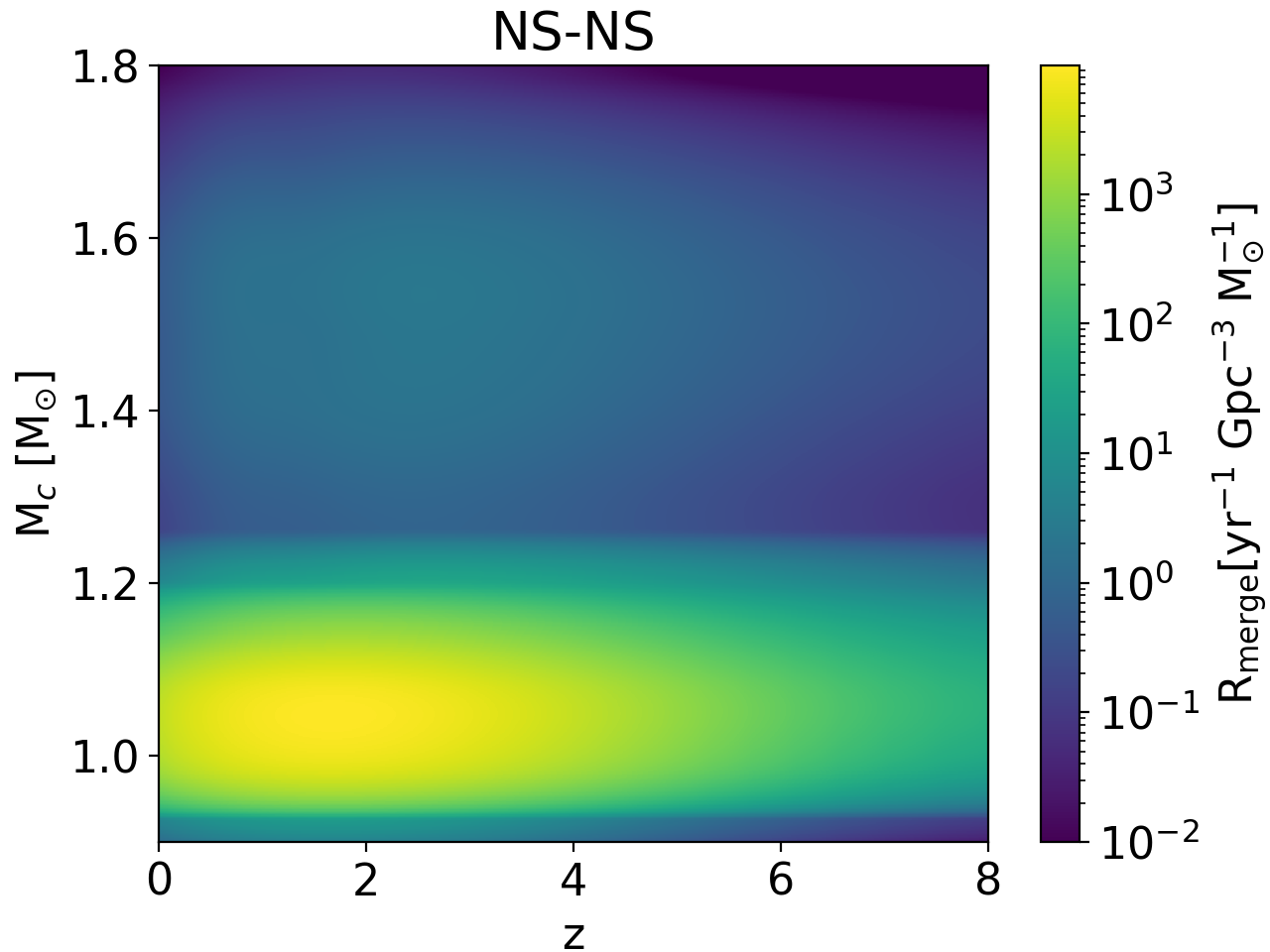}}
\caption{Differential merger rates $d^{2}\dot{N}/ dV \, d\mc$ as a function of redshift and chirp mass for BH-BH (a) and NS-NS (b), obtained combining the output of the STARTRACK population synthesis simulations and observationally driven galactic prescriptions as described in \citep{Boco:2020pgp}. Both the merger rates are normalized to the local values inferred by aLIGO/Virgo \citep{Abbott:2020gyp, LIGOScientific:2021qlt}.} 
\label{fig:merger_rates}
\end{figure}

\begin{figure}
\centering
\subfloat[]
{\includegraphics[width=.49\textwidth]{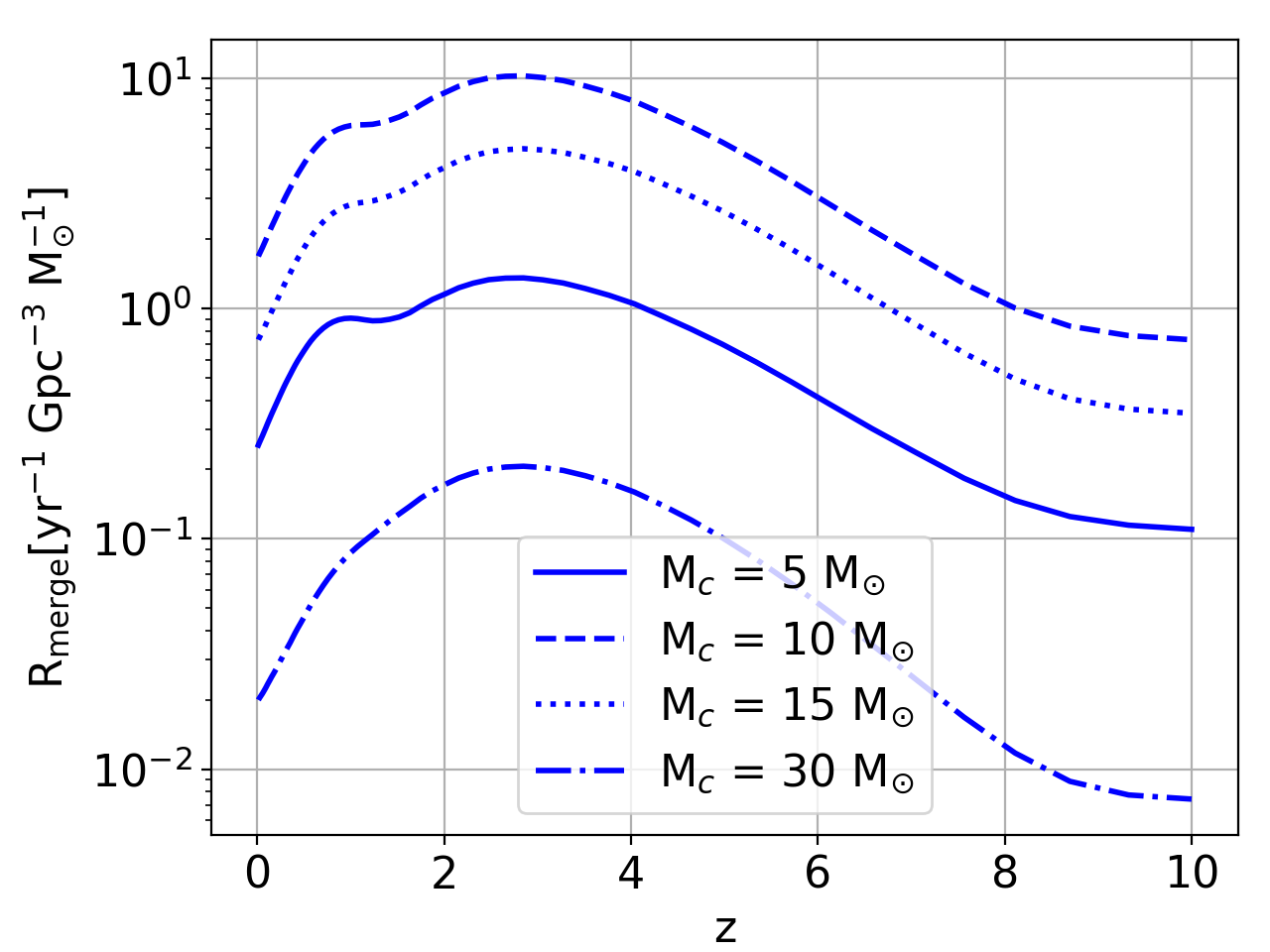}} \ 
\subfloat[]
{\includegraphics[width=.49\textwidth]{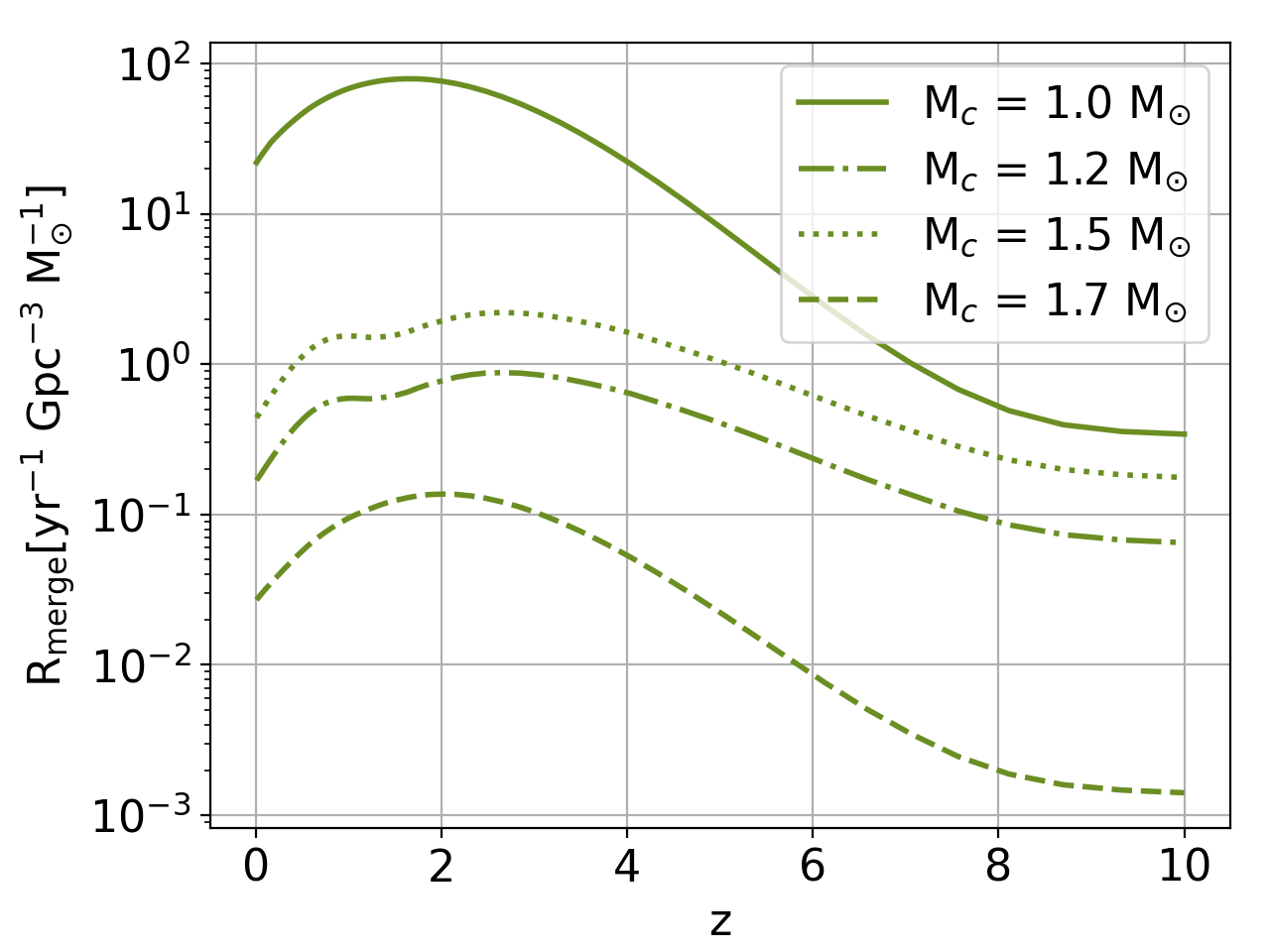}} \\
\subfloat[]
{\includegraphics[width=.49\textwidth]{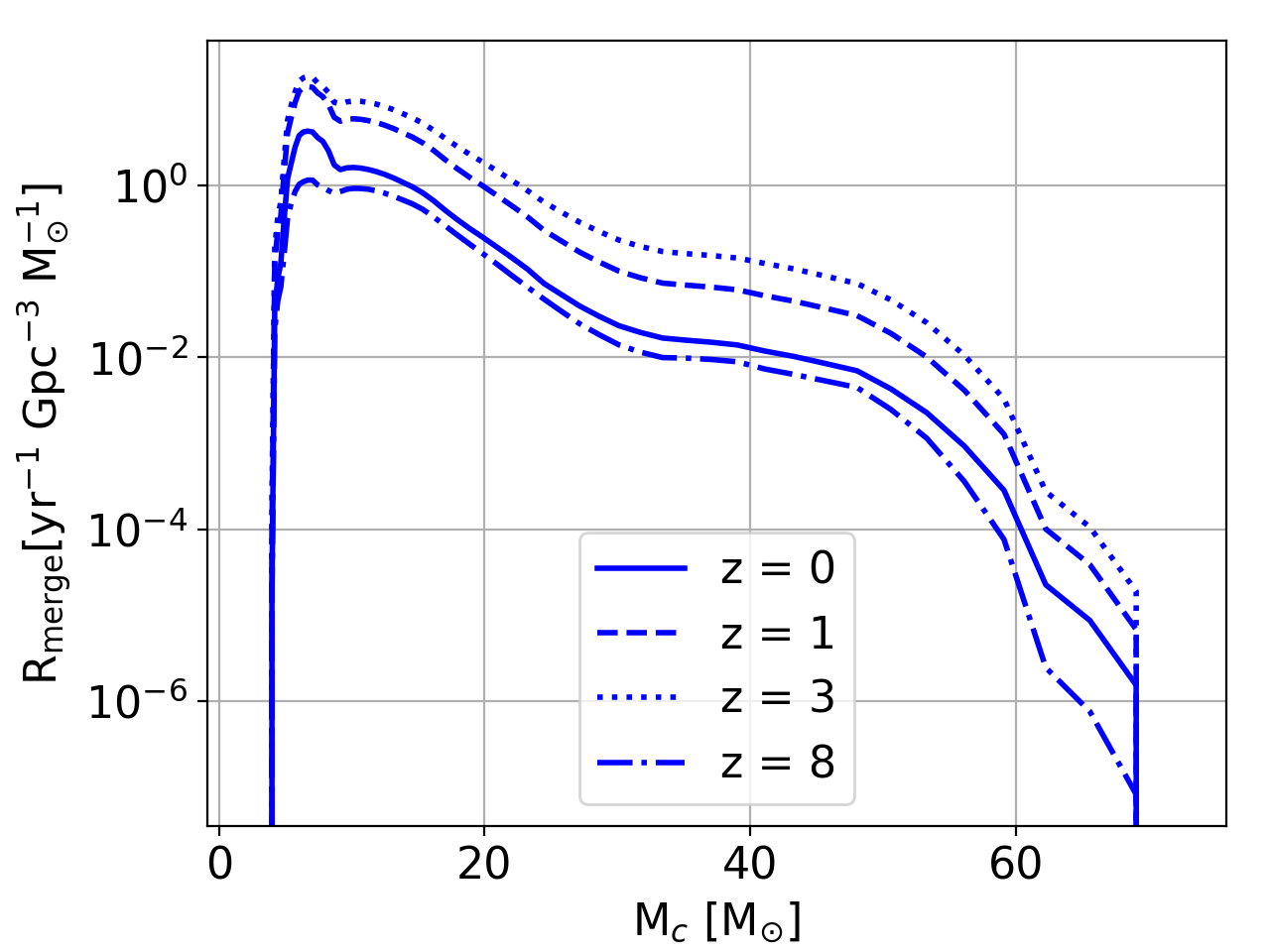}} \ 
\subfloat[]
{\includegraphics[width=.49\textwidth]{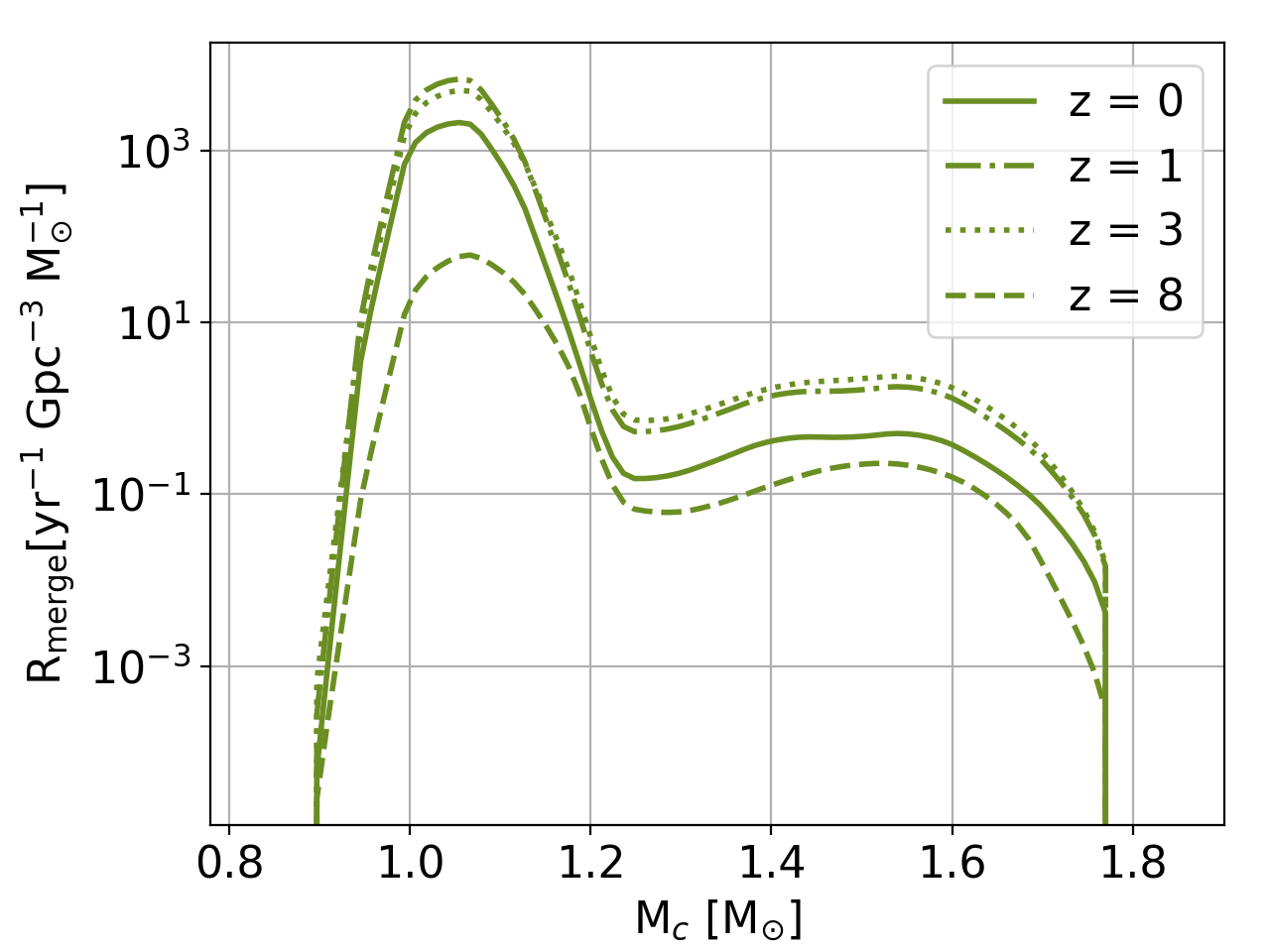}}
\caption{Differential merger rate as a function of redshift at four significant fixed chirp masses for BH-BH (a) and NS-NS (b). Differential merger rate as a function of chirp mass at four different fixed redshifts for BH-BH (c) and NS-NS (d).}
\label{fig:merger_rates_z_mc}
\end{figure}

The resulting merger rates are in good agreement with the recent local determination by the LIGO/Virgo collaboration, as shown in \cite{Boco:2020pgp}. The overall normalization of the merger rates is the result of many different and complex physical processes related to stellar evolution that could in principle depend on the binary type (binary fraction, common envelope development/survival, natal kicks, mass transfers, etc.). In order to reduce the impact of the huge uncertainties in the modeling of these processes, we decide to re-scale our results to match the local merger rates values of the latest LIGO/Virgo estimations, as already done in previous works \citep{Scelfo:2020jyw,Calore:2020bpd,Cao_2017,Li:2018prc}. Incidentally, in this way we also partially reabsorb the effect of adopting the approximated delay time distribution $dp/dt_{d}\propto t_{d}^{-1}$. The original values of the local rates are $32 \,\rm Gpc^{-3}\,yr^{-1}$ for BH-BH, $13 \,\rm Gpc^{-3}\,yr^{-1}$ for BH-NS and $150 \,\rm Gpc^{-3}\,yr^{-1}$ for NS-NS, which are compatible with the aLIGO/Virgo estimations\footnote{Note that we normalize to the mode value given by LIGO/Virgo.} \citep{Abbott:2020gyp, LIGOScientific:2021qlt}: $23.9\substack{+14.9 \\ -8.6}\,\rm Gpc^{-3}\,yr^{-1}$ for BH-BH,  $320\substack{+490 \\ -240}\,\rm Gpc^{-3}\,yr^{-1}$ for NS-NS and $45\substack{+75 \\ -33}\,\rm Gpc^{-3}\,yr^{-1}$ for BH-NS. The re-scaled merger rates as a function of redshift and chirp mass for BH-BH and NS-NS are shown in figure \ref{fig:merger_rates}. The plots in figure \ref{fig:merger_rates_z_mc} show explicitly the redshift (upper panels) and chirp mass (lower panels) dependence of the merger rates. The redshift distribution of the merger rates is highly influenced by the star formation history of the host galaxies: most of the BH-BH events come from $z\sim 2-3$, whereas most of the NS-NS ones come from slightly lower redshifts, $z\sim 2$. The chirp mass dependence, instead, is mainly determined by the stellar prescriptions and the derived DCO mass function, which is largely uncertain in the high mass end, since in that regime different formation channels may enter into play complicating the evolutionary scenario (see e.g. \citep{sicilia_prep}). All in all, the particular features of the merger rates strongly depend on the adopted astrophysical prescriptions: we refer the interested reader to \citep{Boco:2020pgp} for a more detailed treatment.  

Finally, we stress that our method to compute the merger rates presents a twofold benefit: on the one hand the galactic part is completely observational based, not relying on the results of any cosmological simulation or semi-analytic framework; on the other hand, making use of the SFR functions as galaxy statistics, we are able to asses the contribution of galaxies with different properties to the overall DCO merger rate. 

\section{The stochastic Gravitational Wave background}
\label{sec:asgwb}

The spectrum of the SGWB is characterized by the dimensionless energy density parameter
\begin{equation}
\omegagw(\fobs, \eobs ) 
= \dfrac{1}{\rho_{c}} \dfrac{d^{3} \rho_{\rm{gw}}(\fobs, \eobs )}{d \ln \fobs \, d^{2} \omega_{\rm{o}}}
= \dfrac{8 \pi G \fobs}{3H_{0}^{2}c^{2}} \dfrac{d^{3} \rho_{\rm{gw}}(\fobs, \eobs )}{d \fobs \, d^{2} \omega_{\rm{o}}}\,,
\end{equation}
where $\rho_{c} = 3H_{0}^{2}c^{2}/8\pi G$ is the critical density and $\rho_{\rm{gw}}$ is the SGWB energy density at the observed frequency $\fobs$, arriving from a solid angle $\omega_{\rm{o}}$ centred on the observed direction $\eobs$. This quantity can be split into a homogeneous and isotropic term $\baromega (\fobs)$ and a directional dependent GW energy density contrast $\delta \omegagw (\fobs, \eobs)$:

\begin{equation}
\omegagw(\fobs, \eobs) =\dfrac{\baromega (\fobs)}{4 \pi} + \delta \omegagw(\fobs, \eobs)\,.
\end{equation}
The term $\baromega /4\pi$ corresponds to the average GW flux per unit solid angle at the frequency $\fobs$, whereas $\baromega$ is the total GW flux. Since in literature results for isotropic models of the SGWB are usually expressed in terms of the total flux, we are going to express our results in terms of $\baromega$.
The aim of this Section is to explain how we compute the above isotropic and anisotropic terms.

\subsection{Isotropic SGWB}
\label{sec:isgwb}

The isotropic component of the SGWB is constituted by the superposition of all the unresolved GW signals form coalescing compact binaries in galaxies. The intensity of such signals depends on the chirp mass and redshift of the source binary. In fact, the typical waveform of a GW from a binary system is proportional to: 
\begin{equation}
\tilde{h}(f) \propto \mathcal{M}_{z}^{5/6}\, D_{L}^{-1} (z) \, F(f)\,, 
\end{equation}
where $\mathcal{M}_{z}$ is related to the chirp mass $\mc$ through $\mathcal{M}_{z} = \mc (1+z)$, $D_{\rm{L}} (z)$ is the luminosity distance of the source and $F(f)$ is a frequency behavior that depends on the phase of the orbital evolution and is usually described by post-Newtonian approximations. The isotropic energy density of the SGWB is hence directly related to the merger rates of binaries and to the energy spectrum of the GW signal generated by each source, but also to the sensitivity of the observing detector. Therefore, as with increasing detector sensitivity it will be possible to resolve a larger fraction of events out to higher redshift, the total intensity of the SGWB will be reduced. 

All in all, the isotropic energy density of the SGWB is given by the following well known expression (see e.g. \citep{TheLIGOScientific:2016wyq,Abbott:2017xzg, Regimbau:2011rp, Rosado:2011kv, Marassi:2011si, Zhu:2011bd, Zhu:2012xw, Wu:2011ac, KowalskaLeszczynska2015}):
\begin{equation} \label{eq:omega_gw}
\baromega(\fobs)= \dfrac{8 \pi G \fobs} {3 H_{0}^{3} c^{2}} \int dz \int d \mc \;
\dfrac{R_{\rm{merge}}(\mc,z)}{(1+z)\; h(z) } \dfrac{dE}{df} (f_{e}(z) | \mc) \int_{0}^{\bar{\rho}}  d \rho \; P_{\rho} (\rho | \mc, z)\,,
\end{equation}
where $R_{\rm{merge}}= d^{2} \dot{N} / dV d\mc$ is the intrinsic merger rate per unit comoving volume and per unit chirp mass, $h(z) = [ \Omega_{M}(1+z)^{3} + 1 - \Omega_{M}]^{1/2}$ accounts for the dependence of comoving volume on cosmology, $dE/df$ is the energy spectrum of the signal emitted by a single binary evaluated in terms of the source frequency $f_{e} = (1+z) \fobs$ and $P_{\rho}$ is the distribution of signal-to-noise ratio (S/N) for a given detector at given chirp mass and redshift. The rationale behind the previous equation is to compute the energy density of the SGWB summing the contributions of all those GW events whose S/N is below the detection threshold $\bar{\rho} = 8$. Of course, the energy density of the SGWB produced by all the events, resolved and unresolved, can be easily obtained using $\bar{\rho} = \infty$ in equation \eqref{eq:omega_gw}: in this way, the S/N integral is equal to 1 and all the events are taken into account. Following the formalism depicted in \citep{Finn:1995ah,Taylor:2012db}, we compute the distribution of S/N as 
\begin{equation}
P_{\rho} (\rho | \mc, z) = P_{\Theta} (\Theta_{\rho}) \dfrac{\Theta_{\rho}}{\rho}\,,
\end{equation}
in terms of the orientation function
\begin{equation}
\Theta = \dfrac{\rho}{8} \dfrac{D_{L}(z)}{R_{0}}\, \biggl[ \dfrac{1.2 M_{\odot}}{(1+z) \mc}\biggl]^{5/6} \dfrac{1}{\sqrt{\zeta_{\rm{isco}} + \zeta_{\rm{insp}} + \zeta_{\rm{merg}} + \zeta_{\rm{ring}}}}\,,
\end{equation}
and its distribution function
\begin{equation}
P_{\Theta} = 
\begin{cases}
5\Theta (4-\Theta)^{3}/256 \quad & 0<\Theta<4\,, \\
0 \quad & \rm{otherwise} 
\end{cases}\,.
\end{equation}
In the above expressions $R_{0}$ is the detector characteristic distance parameter given by
\begin{equation}
R_{0}^{2} = \dfrac{5}{192} \sqrt{N_{\rm{det}}}\dfrac{M_{\odot}^{2}}{\pi c^{3}} \biggl( \dfrac{3 G}{20}\biggl)^{5/3} x_{7/3}\,,
\end{equation}
where $N_{\rm{det}}$ is the number of co-located detectors and $x_{7/3}$ is the auxiliary quantity
\begin{equation}
x_{7/3} = \int_{0}^{\infty} \dfrac{df}{(\pi M_{\odot})^{1/3} f^{7/3} S(f)}\,,
\end{equation}
with $S(f)$ the noise spectral density. The functions $\zeta_{\rm{isco}}$, $ \zeta_{\rm{insp}}$, $  \zeta_{\rm{merg}} $ and $ \zeta_{\rm{ring}}$ specify the overlap of the waveform with the observational bandwidth during the inspiral, merger and ring-down phases of the event and are given by:
\begin{equation}
\begin{aligned}
& \zeta_{\rm{isco}} = \dfrac{1}{(\pi M_{\odot})^{1/3} \; x_{7/3} } \int_{0}^{2 f_{\rm{isco}} } \dfrac{df}{ S(f)} \dfrac{1}{ f^{7/3}}\,, \\
& \zeta_{\rm{insp}} = \dfrac{1}{(\pi M_{\odot})^{1/3} \; x_{7/3} } \int_{2 f_{\rm{isco}}}^{f_{\rm{merg}} } \dfrac{df}{ S(f)} \dfrac{1}{ f^{7/3}}\,, \\
& \zeta_{\rm{merg}} = \dfrac{1}{(\pi M_{\odot})^{1/3} \; x_{7/3} } \int_{f_{\rm{merg}}}^{f_{\rm{ring}} } \dfrac{df}{ S(f)} \dfrac{1}{ f^{4/3} f_{\rm{merg}}}\,, \\
& \zeta_{\rm{ring}} = \dfrac{1}{(\pi M_{\odot})^{1/3} \; x_{7/3} } \int_{f_{\rm{ring}}}^{f_{\rm{cut}} } \dfrac{df}{ S(f)} \dfrac{1}{ f_{\rm{ring}}^{4/3} f_{\rm{merg}}} \times \biggl[1+ \biggl( \dfrac{f- f_{\rm{ring}}}{\sigma/2}\biggl)^{2} \biggl]^{-2}\,,
\end{aligned} \\
\end{equation}
where $f_{\rm{merge}}, f_{\rm{ring}}, f_{\rm{cut}}$ and $\sigma$ are phenomenological waveform parameters \citep{Ajith:2007kx} and $f_{\rm{isco}}$ is the redshifted frequency at the innermost circular stable orbit:
\begin{equation}
f_{\rm{isco}} \simeq \dfrac{2198}{1+z} \biggl( \dfrac{M_{\rm{bin}}}{M_{\odot}}\biggl)^{-1} \rm{Hz}\,.
\end{equation}
Finally, the energy spectrum emitted by the binary is taken as \citep{Ajith:2007kx, Zhu:2011bd}:
\begin{equation}
\dfrac{dE}{df} \simeq \dfrac{(\pi G)^{2/3} \mc^{5/3}}{3} \times
\begin{cases}
f^{-1/3}  & \quad f<f_{\rm{merg}} \\[5pt]
f_{\rm{merg}}^{-1} \, f^{2/3}  & \quad f_{\rm{merg}} \leq f < f_{\rm{ring}} \\[5pt]
\dfrac{f_{\rm{merg}}^{-1}\, f_{\rm{ring}}^{-4/3} \, f^{2} }{\biggl[ 1+ \biggl( \dfrac{f-f_{\rm{ring}}}{\sigma/2}\biggl)^{2} \; \biggl]^{2}}
& \quad  f_{\rm{ring}} < f \leq f_{\rm{cut}} \\
\end{cases}\,.
\bigskip
\end{equation}   

\subsection{SGWB anisotropies}
\label{sec:sgwba}

Similarly to the isotropic term $\baromega(\fobs)$, also $\delta \omegagw(\fobs, \eobs)$ depends on the local astrophysics at galactic and sub-galactic scales and on the underlying cosmology. However, the anisotropies depend also on the Large Scale Structure (LSS) and its effect on the distribution of sources and on the propagation of GW. Analytic expressions for the anisotropies of the astrophysical SGWB, computed starting from the ingredients listed above, already exist in literature (see e.g. \citep{Cusin:2017fwz,Jenkins:2018nty,Bertacca:2019fnt,Pitrou:2019rjz}).

Since the anisotropies in the total matter density are encoded in the anisotropic component of the SGWB, the latter can be considered a tracer of the former. This argument allows us to exploit the public code  \texttt{CLASS} \citep{lesgourgues2011cosmic, Blas_2011}, without the necessity of modifing it, for computing the angular power spectrum of the SGWB anisotropies. Indeed, the formalism for the computation of the SGWB anisotropy power spectrum is similar to the one used for computing the power spectrum of galaxy number counts. We can therefore use the \texttt{CLASS} routines dedicated to number counts \citep{DiDio:2013bqa} also in our case.
Despite the analogy in the two formalisms, there are indeed deep conceptual differences that have to be dealt with carefully. In fact, the physics of number counts is intrinsically tomographic, whereas the physics of SGWB is somehow blind to the location in redshift space of the sources, since the superposition of all the unresolved events is calculated integrating along $z$. Therefore, in order to adapt the number count formalism to the SGWB, we must reduce the analysis to a single redshift bin that spans from $z=0$ up to a certain $z_{max}$, beyond which the GW contributions are negligible. In this way, all the events in the considered redshift range are integrated together and contribute to the angular power spectrum of the energy overdensity:
\begin{equation}
    C_{\ell} = \dfrac{2}{\pi} \int \dfrac{dk}{k} \, P(k) \, \biggl[ 4\pi \dfrac{\delta\Omega_{\ell}(k)}{\baromega }\biggl]^{2}\,,
\end{equation}
where
\begin{equation}
   \delta \Omega_{\ell}(k) = \int_{0}^{z_{max}} dz \, \dfrac{d\Omega}{dz} W(z,z_{max}/2 ) \, \delta \Omega_{\ell}(k,z)\,. 
\end{equation}
In the previous expressions, $P(k)$ is the primordial matter power spectrum, $d\Omega/dz$ is the density per redshift interval of the tracer, $W(z,z_{i})$ is a window function centred in $z_{max}/2$, which in our case will be a top hat in order to weight all the events equally, and $\delta \Omega_{\ell}(k,z)$ is the relativistic angular fluctuation of the density of the tracer, which is determined by density (den), velocity (vel), lensing (len) and gravity (gr) effects \citep{Bonvin_2011,Challinor_2011,Scelfo:2020jyw}:
\begin{equation} \label{eq:delta_contributions}
\delta \Omega_{\ell}(k,z) = \delta \Omega_{\ell}^{\rm{den}}(k,z) +\delta \Omega_{\ell}^{\rm{vel}}(k,z) + \delta \Omega_{\ell}^{\rm{len}}(k,z) + \delta \Omega_{\ell}^{\rm{gr}}(k,z)\,.
\end{equation}
The relative importance between each of these terms depends on the specific configuration (redshift bins, window functions, etc.) but the main contribution is usually given by the density term. The full expressions of the relativistic effects in the fluctuation $\delta \Omega_{\ell}$ are:
\begin{equation}
\begin{aligned}
\delta \Omega_{\ell}^{\rm{den}}(k,z) =\,& b_{\Omega}\, \delta (k, \tau_{z}) j_{\ell} \\[5pt] 
\delta \Omega_{\ell}^{\rm{vel}}(k,z) =\,& \dfrac{k}{\mathcal{H}} j_{\ell}'' V(k, \tau_{z}) + \biggl[ (f_{\Omega}^{\rm{evo}} - 3) \dfrac{\mathcal{H}}{k} j_{\ell} + \biggl( \dfrac{\mathcal{H}'}{\mathcal{H}^{2}} + \dfrac{2-5s_{\Omega}}{r(z) \mathcal{H}} + 5 s - f_{\Omega}^{\rm{evo}}\biggl)j_{\ell}'  \biggl]V(k, \tau_{z}),  \\[5pt] 
\delta \Omega_{\ell}^{\rm{len}}(k,z) =\,& \ell(\ell +1)\dfrac{2-5s_{\Omega}}{2}\int_{0}^{r(z)} dr \dfrac{r(z)-r}{r(z)r} [\Phi(k, \tau_{z})+\Psi(k, \tau_{z})] j_{\ell}(kr)\,, \\[5pt]
\delta \Omega_{\ell}^{\rm{gr}}(k,z) =\,& \biggl[ \biggl(\dfrac{\mathcal{H}'}{\mathcal{H}^{2}} + \dfrac{2-5s_{\Omega}}{r(z) \mathcal{H}} + 5 s_{\Omega} - f_{\Omega}^{\rm{evo}} +1 \biggl) \Psi(k, \tau_{z}) + (5s_{\Omega}-2)\Phi(k, \tau_{z}) + \mathcal{H}^{-1} \Phi'(k, \tau_{z}) \biggl] j_{\ell} + \\[5pt] 
+& \int_{0}^{r(z)} dr \dfrac{2-5s_{\Omega}}{r(z)} [\Phi(k, \tau)+\Psi(k, \tau)] j_{\ell}(kr) + \\[5pt] 
+& \int_{0}^{r(z)} dr \biggl( \dfrac{\mathcal{H}'}{\mathcal{H}^{2}} + \dfrac{2-5s_{\Omega}}{r(z) \mathcal{H}} + 5 s_{\Omega} - f_{\Omega}^{\rm{evo}} \biggl)[\Phi'(k, \tau)+\Psi'(k, \tau)] j_{\ell}(kr)\,, 
\end{aligned}
\end{equation}
where $b_{\Omega}$ is the bias of the tracer, $s_{\Omega}$ is the magnification bias, $f_{\Omega}^{\rm{evo}}$ is the evolution bias, $r$ is the conformal distance on the light cone, $\tau  = \tau_{0} -r$ is the conformal time, $\tau_{z}  = \tau_{0} -r(z)$, $j_{\ell}, j_{\ell}'$ and $j_{\ell}''$ are the Bessel functions and their derivatives evaluated at $y = kr(z)$ if not explicitly stated, $\mathcal{H}$ is the conformal Hubble parameter, the prime symbol stands for derivatives with respect to conformal time, $\delta$ is the density contrast in the comoving gauge, $V$ is the peculiar velocity, $\Phi$ and $\Psi$ are Bardeen potentials. 

From previous expressions, it can be seen that there are four main ingredients that are needed to fully characterize our tracer: the \textit{redshift distribution}   $d\Omega/dz$, i.e. the source number density per redshift interval, the \textit{bias}  $b_{\Omega}$, which quantifies the mismatch between the distribution of the matter and of the tracer, the \textit{magnification bias}  $s_{\Omega}$, which quantifies the change in the observed surface density of sources of tracer induced by gravitational lensing and the \textit{evolution bias}  $f_{\Omega}^{\rm{evo}}$, which reflects the fact that the number of elements of a tracer is not necessarily conserved in redshift due to the possible formation of new objects. All these quantities for the SGWB will be first formally derived and then numerically computed in Sub-section \ref{sec:ctsgwbaatom}.

\section{Results}
\label{sec:results}

In this Section, we report the most relevant results of our work. First of all, we discuss the features of the isotropic energy density of the SGWB originated by our astrophysical sources. Secondly, we present a full characterization of the energy density of the SGWB produced by coalescing binaries as a tracer of the distribution of galaxies and consequently of the total matter density. We then show our results for the power spectrum of the SGWB anisotropies. Finally, we develop a methodology for producing a full sky SWGB anisotropy map. All our analyses are performed for two different detectors: advanced LIGO/Virgo \citep{TheLIGOScientific:2016wyq,PhysRevD.102.062003} and Einstein Telescope \citep{Sathyaprakash:2012jk,Regimbau:2012ir,Maggiore:2019uih,Hild:2010id,Mentasti:2020yyd}.

\subsection{Energy density of the astrophysical SGWB}
\label{sec:edotasgwb}

\begin{figure}
\centering
\includegraphics[width=.6\textwidth]{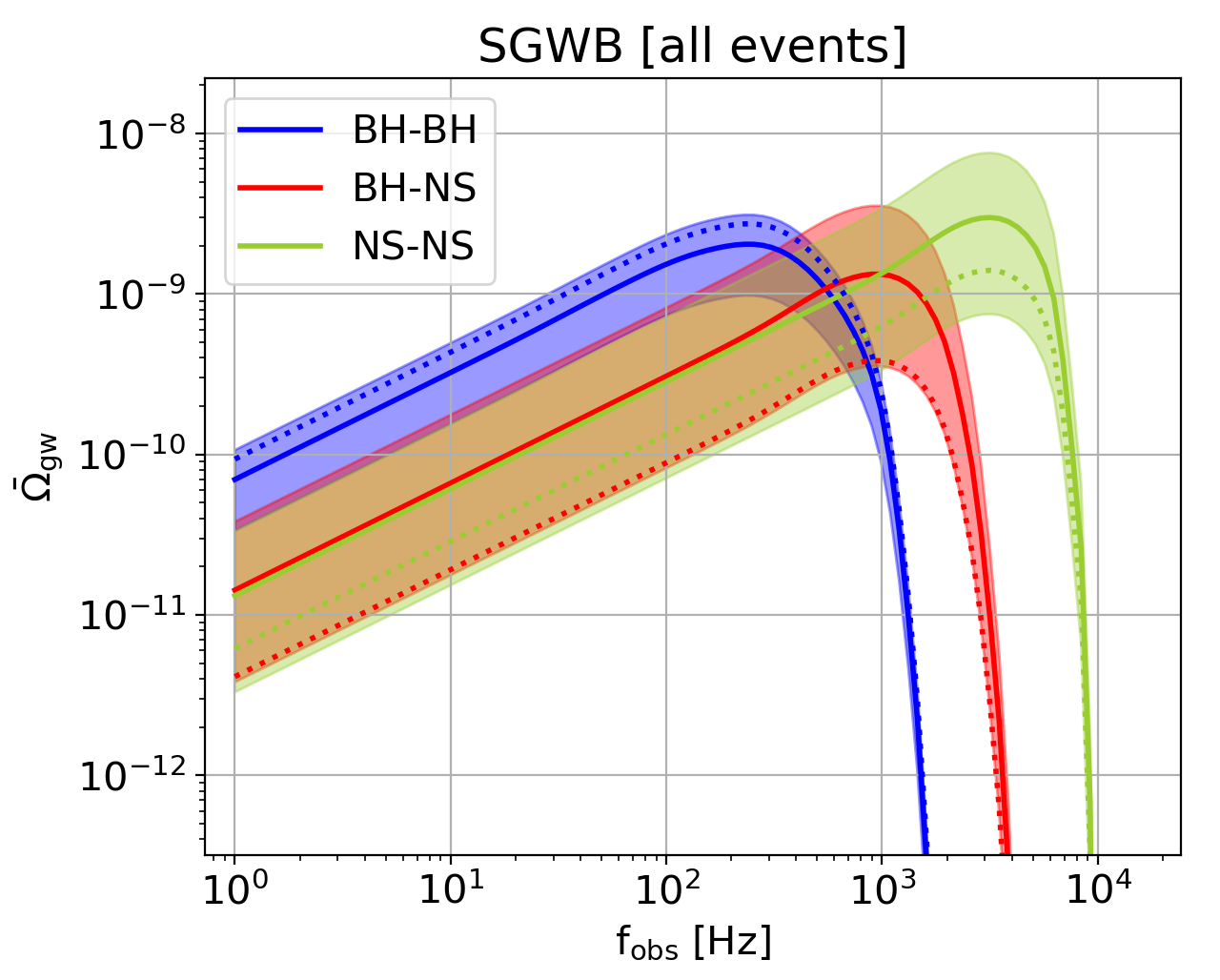}
\caption{Energy density parameter of the SGWB constituted by all the events, resolved and unresolved, as a function of the observed frequency. The curves have been evaluated within the astrophysical set-up described in section \ref{sec:astro}, using equation \eqref{eq:omega_gw} with with $\bar{\rho} = \infty$. Different colors are associated to the SGWB produced different type of merging binaries: BH-BH (blue), BH-NS (red) and NS-NS (green). The solid curves are obtained using the merger rates normalized to the local rate inferred by aLIGO/Virgo \citep{Abbott:2020gyp, LIGOScientific:2021qlt}, whereas the dotted curves are obtained with the original rates. The shaded areas mark the range of amplitudes defined by the error bars of the aLIGO/Virgo estimations. Note that the curves evaluated with the original rates lie within the shaded areas at each frequency and for all binary types.} 
\label{fig:omega_gw_all}
\end{figure}

\begin{figure}
\centering
\includegraphics[width=\textwidth]{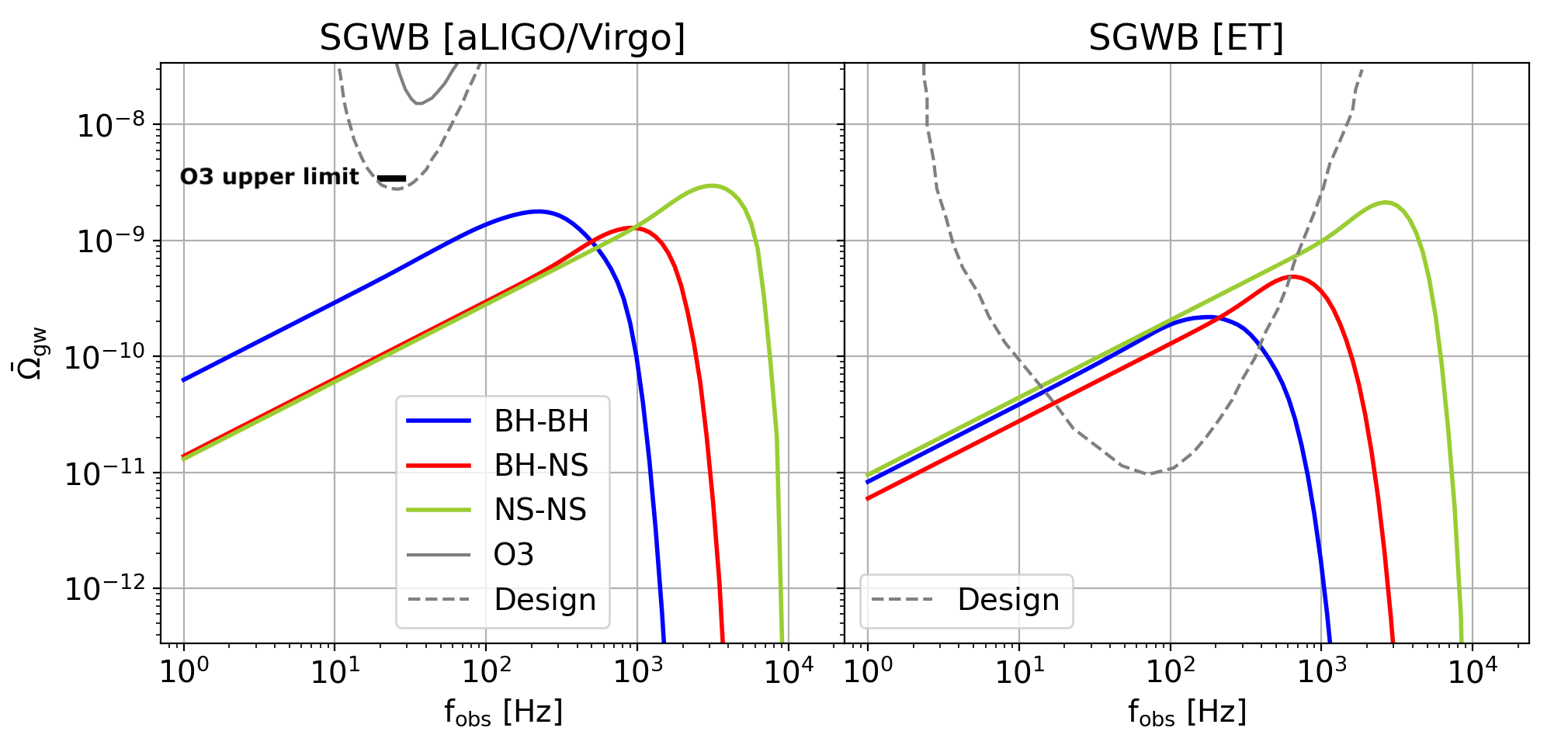}
\caption{Energy density parameter of the SGWB as a function of the observed frequency for aLIGO/Virgo (left panel) and ET (right panel) computed using equation \eqref{eq:omega_gw} within the astrophysical set-up described in section \ref{sec:astro}, using $\bar{\rho} = 8$. The color code is the same as in figure \ref{fig:omega_gw_all}. The dashed grey curves represent the power-law integrated sensitivity curves for the design configuration of the considered detector. In the aLIGO/Virgo panel, we also show the O3 sensitivity (solid grey curve) and the current upper limit on the isotropic SGWB: $\omegagw \leq 3.4 \times 10^{-9}$ at 25 Hz for a power-law background with a spectral index of 2/3, expected for compact binary coalescences \citep{Abbott:2021upperlimits}.}
\label{fig:omega_gw}
\end{figure}

In this Sub-section we present our results regarding the frequency spectrum $\baromega (\fobs)$ of the isotropic energy density of the SGWB. 
We compute the energy density for the three types of DCOs, BH-BH, BH-NS and NS-NS, and we specify our results to the case where all the events, resolved and unresolved, are taken into account (figure \ref{fig:omega_gw_all}) and to the cases where we subtract the events resolved by aLIGO/Virgo and ET (figure \ref{fig:omega_gw}). All the curves have been evaluated in the astrophysical set-up described in section \ref{sec:astro}, using equation \eqref{eq:omega_gw}. In particular, we consider the merger rates normalized to the local values inferred by aLIGO/Virgo. For completeness, in figure \ref{fig:omega_gw_all} we also plot the curves obtained with the original (non-normalized) rates and we show that they lie within the range of amplitudes defined by the error bars of the aLIGO/Virgo estimations.

All the curves in figures \ref{fig:omega_gw_all} and \ref{fig:omega_gw} display the typical shape expected for a SGWB given by the superposition of signals produced by coalescing DCOs. At low frequencies, the dominating contribution comes from the inspiral phase and the energy density is proportional to $f^{2/3}$. As the frequency increases, the contribution from the merger phase becomes more and more important and the power-law slows down until eventually the energy density starts to decrease. The frequency at which this drop occurs is related to the typical mass of the considered DCOs: the lower is the chirp mass of the binary, the higher is the frequency at which the merger occurs. For this reason, the energy density of the SGWB produced by coalescing BH-BH binaries peaks at lower frequencies than the energy density of the SGWB produced by coalescing NS-NS. Finally, very high frequencies are associated to the ringdown phase and are characterized by an exponential drop. The great effect of different astrophysical prescriptions on the energy density \citep{KowalskaLeszczynska2015,Cusin:2019jpv} makes it difficult to do an accurate comparison with previous results. However, we can say that both shape and amplitude of our detector-independent spectra in figure \ref{fig:omega_gw_all} are compatible with the results presented in e.g. \citep{Regimbau:2011rp,Rosado:2011kv,Marassi:2011si}.

The ratio of the amplitudes of the backgrounds coming from different types of coalescing binaries depends on the merger rates, on the typical strength of the signals and on the number of events resolved by the detector. In our astrophysical framework, NS-NS binary systems have the largest merger rate density, followed in order by BH-NS and NS-NS. On the other hand, binary systems with higher chirp mass produce a louder GW signal, so that BH-BH events are the most intense, followed by BH-NS and NS-NS. Without considering the effect of the detector, these two effects combine together such that the BH-BH systems produce the SGWB with the greatest amplitude, followed by NS-NS and BH-NS. For aLIGO/Virgo this order is preserved since the number of resolved events is moderately low. The situation is quite different for ET, which is expected to detect most of the BH-BH systems up to $z\sim 2$ and most of the NS-NS systems up to $z\sim 1$, as we will discuss in more detail in the next Sub-section. Accordingly, for ET the backgrounds produced by BH-BH and BH-NS events are slightly lower than the one produced by NS-NS ones. The higher sensitivity of ET is also responsible of the fact that the energy density is lower for ET than for aLIGO/Virgo: since ET will resolve a larger number of events, there will be less unresolved events to produce the SGWB. 

Finally, we compare our predictions with the sensitivities of the considered detectors. The dashed grey curves in figure \ref{fig:omega_gw} represent the power-law integrated sensitivity curves \citep{Thrane:2013oya} for aLIGO/Virgo and ET at design configuration. As a third generation detector, ET will be more sensitive and it will span a larger frequency range than second generation detectors like aLIGO/Virgo. 
The expected signal for aLIGO/Virgo is one order of magnitude below the current upper limit obtained during the first half of O3, which is equal to $\omegagw \leq 3.4 \times 10^{-9}$ at 25 Hz for a power-law background with a spectral index of 2/3. On the other hand, the exquisite sensitivity of ET will probably allow to detect the astrophysical SGWB, despite the reduced amplitude of the signal due to the large number of resolved sources.
Since both aLIGO/Virgo and ET are more sensitive to SGWBs between a few tens and a few hundreds of Hz, in the rest of the paper we are going to work at an intermediate reference frequency of 65 Hz.

\subsection{Characterizing the SGWB as a tracer of matter}
\label{sec:ctsgwbaatom}

As we stressed already, in the present work we deal with the SGWB given by the superposition of the GW signals produced by merging binaries. Indeed, these GW signals originate in galaxies and therefore trace their distribution. Since galaxies trace the total matter distribution, the energy density of the SGWB itself is a tracer of matter. In this Sub-section, we characterize the astrophysical SGWB as a tracer of the total matter density in order to be able to compute the power spectrum of the anisotropies with \texttt{CLASS}. As already pointed out in Section \ref{sec:asgwb}, the four physical quantities that characterize a tracer are its redshift distribution, its bias and magnification bias and its evolution bias. In the following, we derive from scratch all these quantities for the energy density of the SGWB and we compute them numerically in the astrophysical framework we described in Section \ref{sec:astro}. 

\begin{figure}
\centering
\includegraphics[width=.6\textwidth]{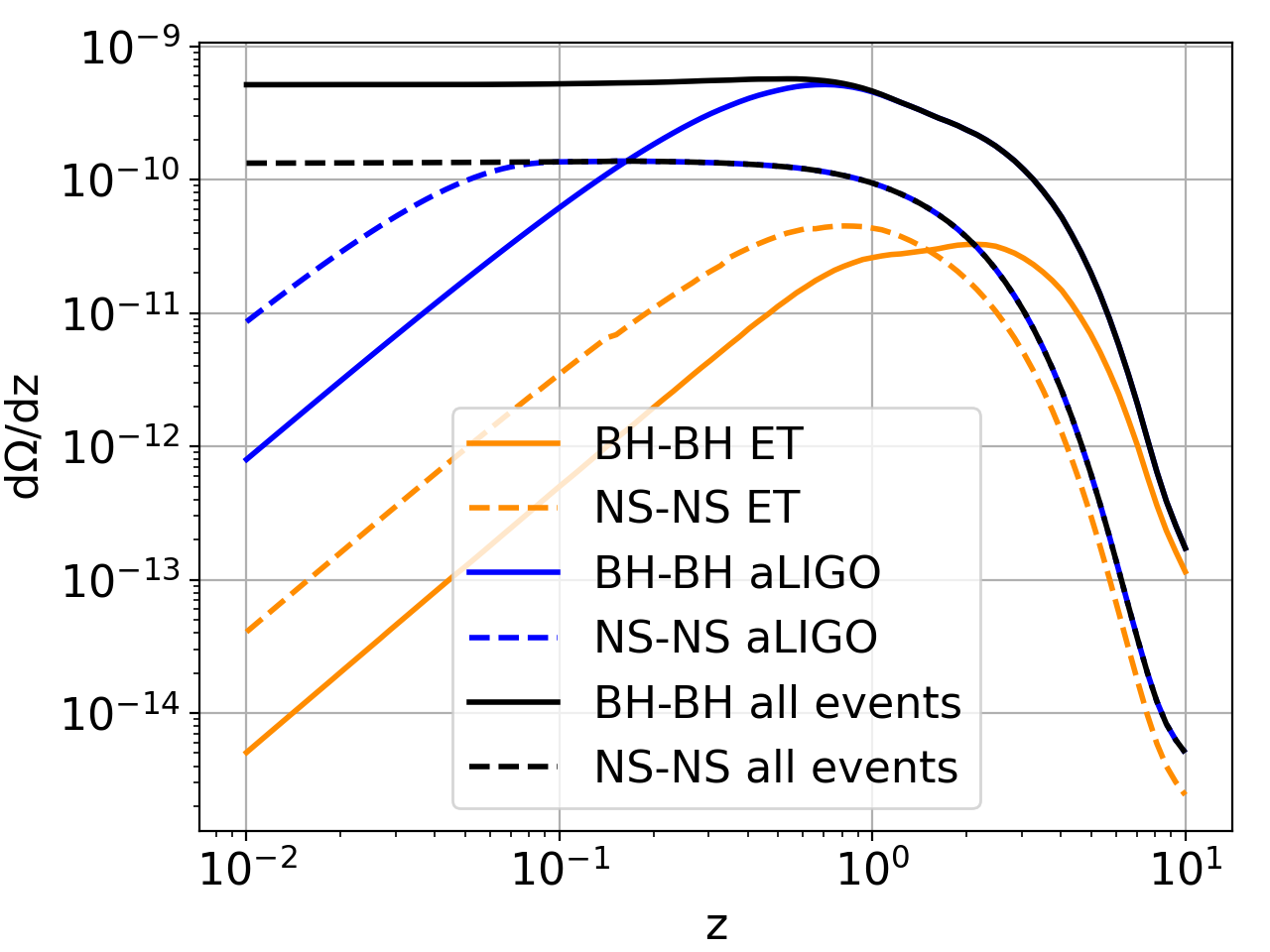}
\caption{Distributions $d\baromega / dz$ for BH-BH (solid) and NS-NS (dashed) merging events at 65 Hz. The black curves are obtained considering the contribution to the SGWB of all the events, whereas the blue and orange distributions are computed subtracting the contributions of the events resolved by aLIGO/Virgo and ET, respectively.}
\label{fig:domega_dz}
\end{figure}

By definition, we can express the isotropic part of the energy density as:
\begin{equation}
    \baromega \equiv \int dz \, \dfrac{d \baromega }{dz}\,. 
\end{equation}
Therefore, the redshift distribution can be easily computed from \eqref{eq:omega_gw} not integrating over redshift:
\begin{equation} \label{eq:domega_dz} 
\begin{split}
\dfrac{d\baromega}{dz} (z,f_{\rm{obs}}) = \, \dfrac{8 \pi G f_{\rm{obs}} }{3 H_{0}^{3} c^{2}} \dfrac{1}{(1+z)\; h(z) } & \int d \mc \;
\dfrac{d^{2} \dot{N}}{d\mc dV } (\mc,z) \dfrac{dE}{df} (f | \mc) \times \\[5pt]
& \times \int_{0}^{\bar{\rho}}  d \rho \; P_{\rho} (\rho | \mc, z)\,.
\end{split} 
\end{equation} 
The redshift distribution is also a function of the observed frequency, but in the following we focus on a single frequency, 65 Hz, at which both aLIGO/Virgo and ET are more sensitive to SGWB (see sensitivity curves in figure \ref{fig:omega_gw}). In figure \ref{fig:domega_dz} we plot $d\baromega/dz$ as a function of redshift at the reference frequency. We show both the contributions to the SGWB from all the merging events ($\bar{\rho} = \infty$ in the previous equation) and subtracting the resolved events by aLIGO/Virgo and ET respectively. The distribution $d\baromega/dz$ for all the events is almost flat up to $z\lesssim 1$ due to a compensation between the increasing number of mergers and the dilution of the GW flux, it start decreasing at $1<z<2$ since merging rates flatten at such redshifts, not compensating anymore for flux dilution, and then decreases rapidly due to the increasing rarity of merging binaries at high redshifts. The curves for aLIGO/Virgo and ET, on the other hand, reflect the fact that GW detectors resolve many of the nearby events (especially the brighter BH-BH ones), which hence do not contribute to the background. For this reason, the redshift distributions for aLIGO/Virgo and ET decrease also at low redshifts, so that intermediate redshift events contribute the most to the energy density of the SGWB. This effect is more important for ET, which has a higher sensitivity than aLIGO/Virgo and is expected to resolve the majority of BH-BH events up to $z\lesssim 2$ and many NS-NS events up to $z\lesssim 1$. Consequently, the bulk of the energy density will come from events located at $z\sim 1-2$ for ET.

\begin{figure} 
\centering
\includegraphics[width=.6\textwidth]{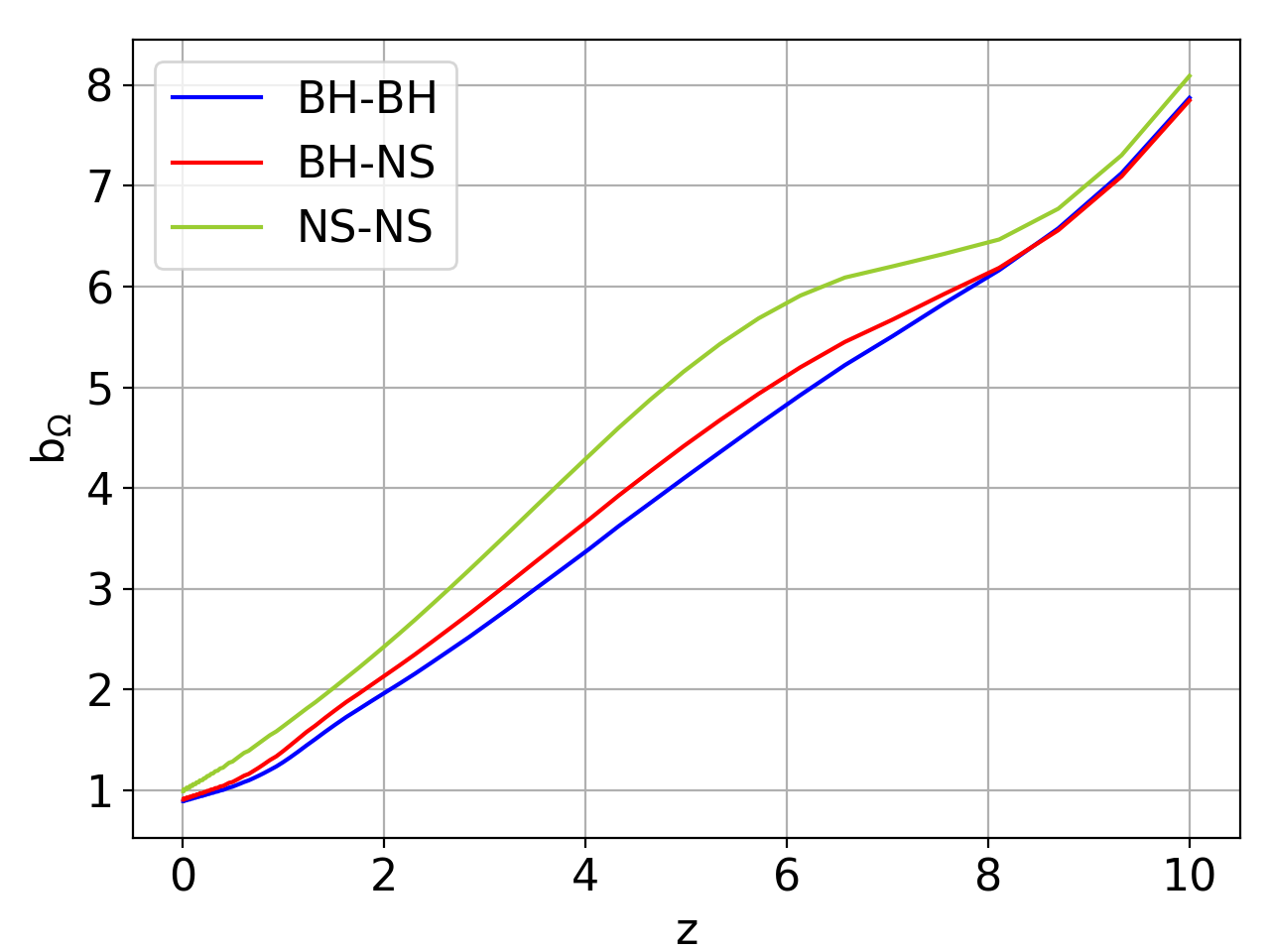}
\caption{Bias of the SGWB energy density for the three types of DCO mergers (BH-BH, BH-NS and NS-NS) as a function of redshift at 65 Hz for ET.}
\label{fig:b_omega}
\end{figure}

The bias $b_{\Omega}$ of the SGWB energy density quantifies the mismatch between the distribution of our tracer, the SGWB, and the total matter density. The background we are considering is produced by GW signals emitted by merging DCOs in galaxies and hence its bias is directly related to the galaxy bias. 
In particular, we make use of the bias $b(z, \psi)$, associated to a galaxy at a given redshift with given star formation rate $\psi$.
Following the procedure depicted in \citep{Aversa:2015bya}, we associate the SFR of each galaxy to the mass of the hosting dark matter halo through an abundance matching technique and then we assign to a galaxy with given SFR the bias of the corresponding halo. The abundance matching is a standard method to derive a monotonic relationship between the galaxy and the halo properties by matching the corresponding number densities in the following way:
\begin{equation}
    \int_{\log_{10}\psi}^{\infty} d \log_{10}\psi' \dfrac{d^{2}N}{d \log_{10}\psi' \, dV} = 
    \int_{-\infty}^{\infty} d \log_{10}M'_{H} \dfrac{d^{2}N}{d \log_{10}M'_{H} \, dV} \, \dfrac{1}{2} \, 
    erfc \biggl[ \dfrac{\log_{10}(M_{H}(\psi))/M'_{H}}{\sqrt{2} \Tilde{\sigma}} \biggl]\,,
\end{equation}
where $d^{2}N/d\log_{10}\psi' / dV$ is the star formation rate function and $d^{2}N/d \log_{10}M'_{H} / dV$ is the galaxy halo mass function, i.e. the mass function of halos that are hosting one individual galaxy (see Appendix A of \citep{Aversa:2015bya} for further details). $M_{H}(\psi)$ is the relation we are looking for and $\Tilde{\sigma} \equiv \sigma d \log_{10}M_{H} / d \log_{10} \psi $ is the scatter around that relation (we set $\sigma_{\log_{10}\psi} \simeq 0.15$). Once $M(\psi)$ is determined, we assign to each galaxy the bias corresponding to the halo associated with its SFR, i.e $b(z,\psi)= b(z,M_{H}(z,\psi))$, where $b(z,M_{H})$ is computed as in \citep{Sheth:1999su} and approximated as in \citep{Lapi:2014ija}.
In order to assign a redshift dependent bias to the SGWB energy density, we weight the galaxy bias $b(z, \psi)$ through the energy density per unit redshift and SFR, which keeps into account the contribution of GW produced in galaxies with different SFRs to the signal. The energy density per unit redshift and SFR can be computed in the same way as the redshift distribution (equation \eqref{fig:domega_dz}), using the merger rate per unit volume, chirp mass and SFR, $d^{2} \dot{N} / d\mc \, dV \, d \log_{10}\psi$, instead of the merger rate per unit volume and chirp mass. Therefore, we define the bias for the SGWB energy density through the following expression:
\begin{equation} \label{eq:b_omega}
    b_{\Omega}(z,f_{\rm{obs}}) 
    \equiv \dfrac{\int d \log_{10}\psi \dfrac{d^{2} \baromega }{dz d \log_{10}\psi}(z,\psi,f_{\rm{obs}}) \; b(z,\psi)}
    {\int d \log_{10}\psi \dfrac{d^{2} \baromega}{dz d \log_{10}\psi}(z,\psi,f_{\rm{obs}})}\,.
\end{equation}
In figure \ref{fig:b_omega} we show the bias of the SGWB energy density for the three types of DCO mergers (BH-BH, BH-NS and NS-NS) as a function of redshift at 65 Hz for the ET. Since it mainly reflects the behavior of the galaxy bias, the SGWB bias increases with redshift, independently on the type of merger. The bias for BH-BH and BH-NS is smaller than the one for NS-NS because BHs need a lower metallicity environment to form. A lower metallicity environment, in turn, requires a lower SFR. Since low SFR galaxies have a smaller bias at all redshifts, the SGWB produced by DCOs containing at least a BH comes from galaxies with smaller bias on average.

\begin{figure} 
\centering
\includegraphics[width=.6\textwidth]{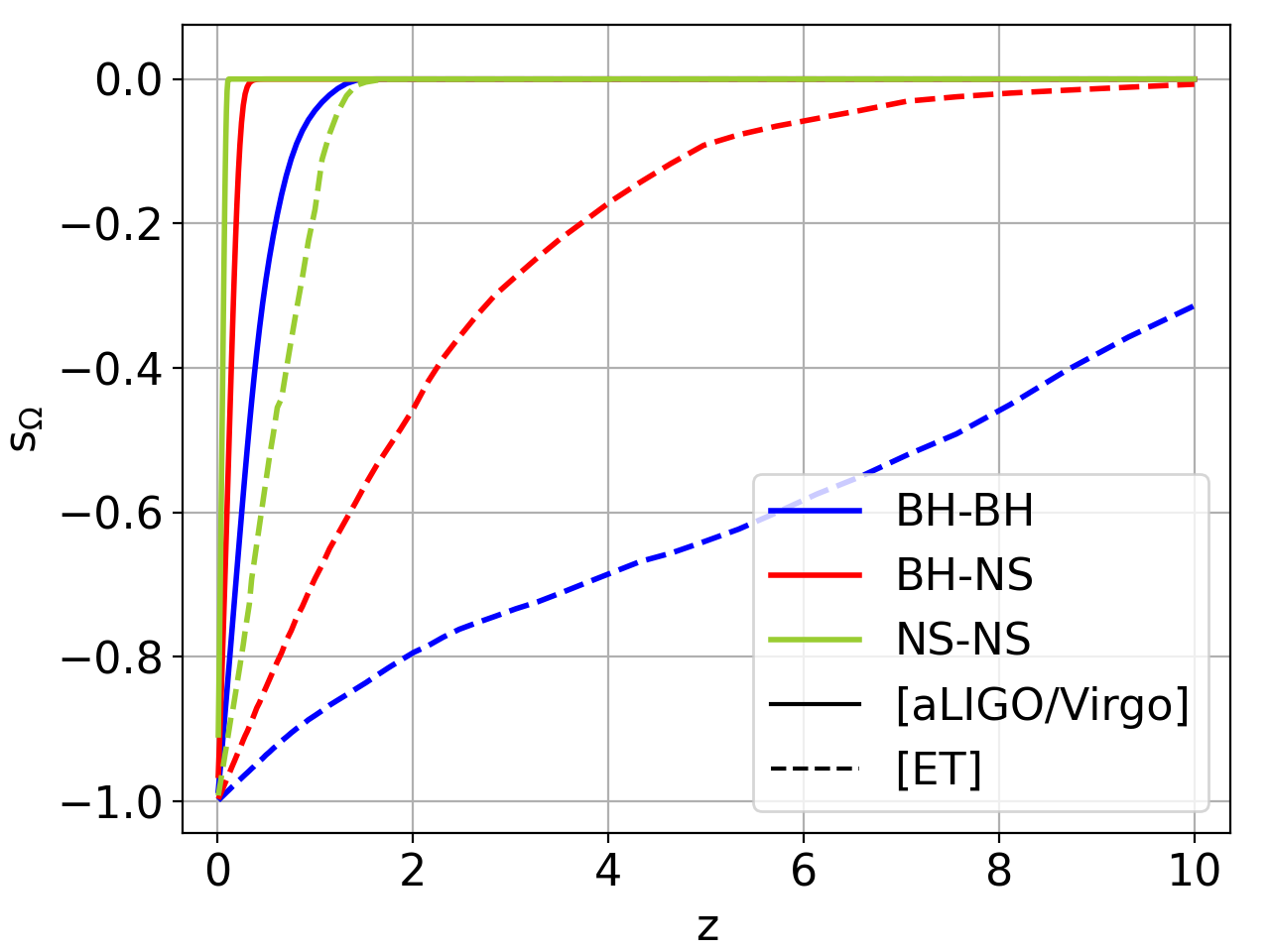}
\caption{Magnification bias of the SGWB energy density for the three types of DCO mergers (BH-BH, BH-NS and NS-NS) as a function of redshift at 65 Hz for aLIGO/Virgo (solid lines) and ET (dashed lines).}
\label{fig:s_omega}
\end{figure}

Let’s now move to the analysis of the effect of gravitational lensing on the SGWB. A gravitational lens with magnification $\mu$ enhances the signal-to-noise ratio $\rho$ of a given GW signal of a factor $\sqrt{\mu}$, without changing the observed waveform (due to the achromaticity of lensing in the geometrical-optics limit). Weak lensing can affect the SGWB in two ways: on the one hand, some of the events that make up the signal are boosted; on the other hand, the associated dilution of the volume reduces the received GW flux. The overall effect is encoded in the magnification bias. Adapting the general derivation that is found in the appendix of reference \citep{Hui_2007} to the case of SGWB, we define the magnification bias of the SGWB energy density through
\begin{equation}
    \dfrac{d\baromega^{\rm{lensed}}}{dz} \equiv \dfrac{d\baromega}{dz} \bigl[1+ \kappa(5s_{\Omega} -2)], 
\end{equation}
where $\kappa$ is the lensing convergence. After some manipulation, we obtain the explicit expression 
\begin{equation} \label{eq:s_omega}
    s_{\Omega, \bar{\rho}}(z,f_{\rm{obs}}) = - \dfrac{1}{5} \dfrac{d \log_{10} \bigl( \frac{d \baromega (f_{\rm{obs}},z, <\rho)}{dz}\bigl)}{d \log_{10} \rho} \Biggl| _{\rho = \bar{\rho}}\,.
\end{equation}
The previous expression for the magnification bias is similar to the one derived in \citep{Scelfo:2018sny,Scelfo:2020jyw} for resolved GW events: a crucial difference is that we are now considering only the GW events that are below the detection threshold $\bar{\rho}$.
Actually, when we compute the SGWB considering all the events, resolved and unresolved, the two effects of weak lensing - growth of energy density due to magnification and dilution of flux - balance each other. Therefore, weak lensing does not affect the anisotropies of the SGWB in this case, as already shown in \citep{Bertacca:2019fnt}. For this reason,  we set the magnification bias to 0.4, i.e.  the value for which the two competing effects cancel out. 
The situation in different when we take into account the detector sensitivity by subtracting the resolved GW events from the energy budget. In this case, lensing magnification boosts some events above the detection threshold, so that they are resolved by the detector and do not contribute to the energy density. This is reflected by a negative magnification bias, as it can be seen in figure \ref{fig:s_omega}, where we plot $s_{\Omega}(z)$ for aLIGO/Virgo and ET. The magnification bias is different from zero where there is a considerable amount of events with a signal-to-noise ratio next to the detection threshold. Accordingly, for BH-BH it goes to zero at higher redshifts than for BH-NS and  NS-NS, since BH-BH events are more energetic and can be resolved at longer distances. For the same reason, the magnification bias for the extremely sensitive ET is lower than zero up to very high redshifts, especially for BH-BH. For what concerns strong lensing, instead, we verified that the effect of lenses with magnification $\mu >2$ on the SGWB energy density is negligible.

Finally, the evolution bias for the SGWB can be written as:
\begin{equation}
f^{\rm{evo}}_{\Omega} = \dfrac{d \ln \bigl( \frac{d\baromega}{dz d\omega} \bigl)}{d \ln a}\,, 
\end{equation}
where $a$ is the scale factor and $\omega$ is the solid angle. The evolution bias appears only in sub-leading contributions, since it is just present in the non-dominant part of the velocity term and in the gravity term of equation \eqref{eq:delta_contributions}.

\subsection{Angular power spectrum of SGWB anisotropies}
\label{sec:apsosgwba}

The distribution in the sky of the SGWB from astrophysical sources has been studied in earlier works \citep{Cusin:2018rsq,Cusin:2019jpv,Jenkins:2018uac,Jenkins:2018kxc,Jenkins:2019nks}. The differences in the signals calculated in those works have been noticed and discussed recently \citep{Cusin:2018ump,Jenkins:2019cau}, suggesting that they may be due to different choices in the astrophysical modeling, and a different treatment of density perturbations and galaxy clustering. Our purpose here is to present the predicted signal deriving from the framework described in previous sections, featuring new descriptions of the stellar and galactic physics. 
We now discuss our results and analyze the phenomenology of our predictions, relating those to the modeling framework we just mentioned.

We compute the angular power spectrum of the SGWB anisotropies in the redshift bin $z \in [0,8]$, using a tophat window function in order to weight all the events equally. We perform the calculation exploiting the \texttt{CLASS} public code. A relevant feature of \texttt{CLASS} is that it allows to insert only one value of bias and magnification bias per redshift bin. This is indeed a limitation, since both bias and magnification bias vary significantly with redshift, as it can be seen in figures \ref{fig:b_omega} and \ref{fig:s_omega} respectively. To overcome this problem, we used a weighted mean of the two functions in the considered redshift interval, where the weight is given by the redshift distribution of the SGWB energy density:
\begin{equation}
    \langle b_{\Omega} \rangle = \dfrac{\int dz \, \dfrac{d \baromega}{dz} \, b_{\Omega}}{\int dz \, \dfrac{d\baromega}{dz}}
    \quad \rm{and} \quad
    \langle s_{\Omega} \rangle = \dfrac{\int dz \, \dfrac{d \baromega}{dz} \, s_{\Omega}}{\int dz \, \dfrac{d\baromega}{dz}}
\end{equation}
Of course, given the strong redshift dependence of bias and magnification bias, this is an approximation that should be improved in the future by modifying the \texttt{CLASS} code in order to obtain more precise results.

\begin{figure} 
\centering
\includegraphics[width=.6\textwidth]{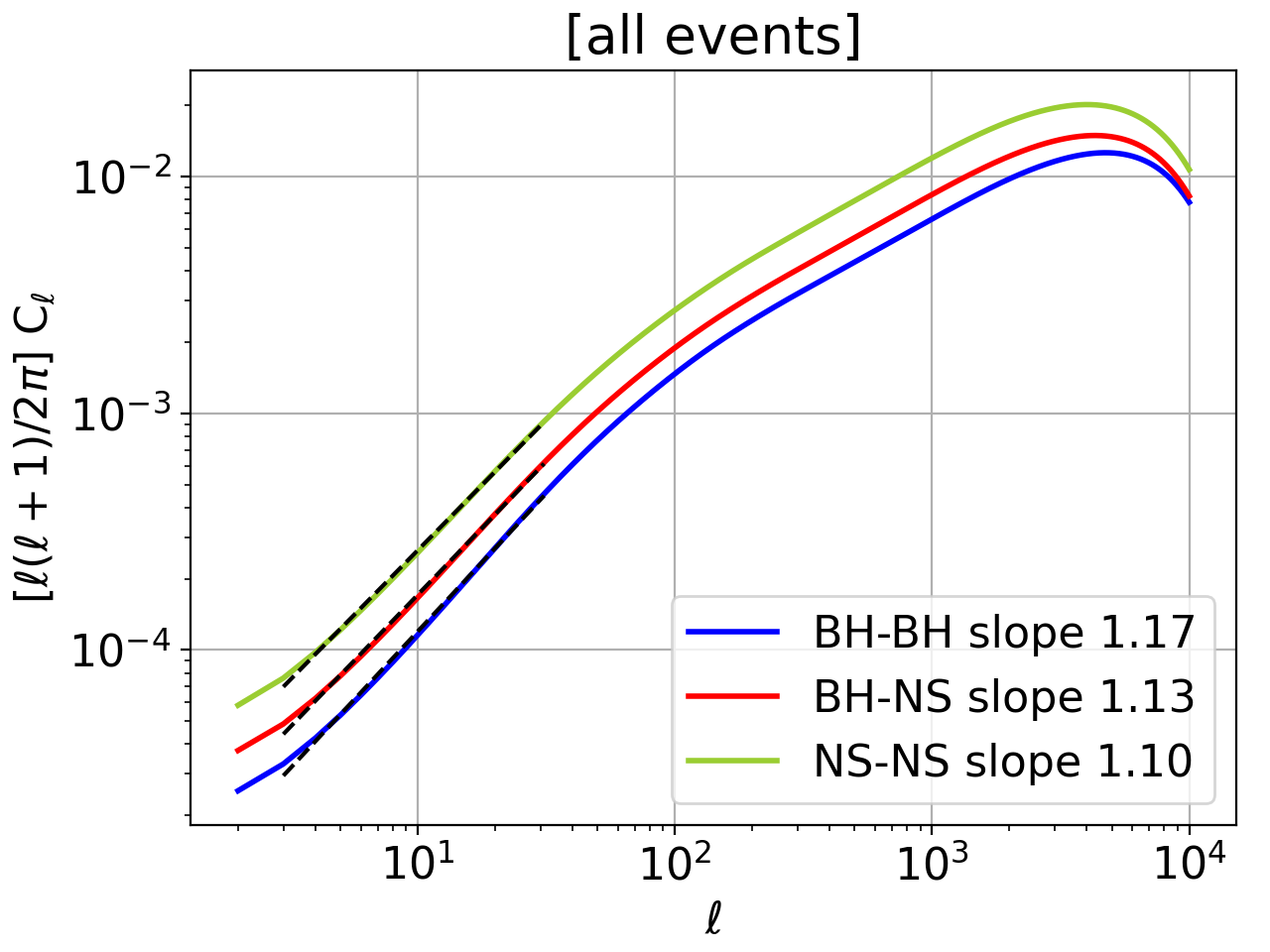}
\caption{Angular power spectrum of the anisotropies of the SGWB produced by BH-BH, BH-NS and NS-NS at 65 Hz. Here, the SGWB is obtained integrating the signals of all the GW events, resolved and unresolved. The black dotted lines represent the power-law fit at large scales, whose slopes are also reported.}
\label{fig:Cls_all_events}
\end{figure}

As a first result, we computed the power spectrum of the anisotropies of the SGWB produced by the superposition of all GW events, resolved and unresolved. As we already stressed, the SGWB measured by a real detector is given by the superposition of the unresolved events only. Nevertheless, studying the anisotropies of the background produced by all events is interesting at least for two reasons: first of all, it is a detector-independent estimate of the expected amplitude of the signal, and second, it is useful in order to compare our results with the other quantitative predictions already present in literature. The power spectrum of the anisotropies for BH-BH, BH-NS and NS-NS at 65 Hz are shown in figure \ref{fig:Cls_all_events}. Independently on the type of source, the curves behave as a power-law at large scales and bend at smaller scales, reaching a peak at $l\sim 5 \times 10^3$. The slope of the power-law is close to 1, which means that $C_{\ell} \propto 1/\ell$ approximately, as it was also found in \citep{Cusin:2018rsq,Cusin:2019jpv}.

\begin{figure}
\centering
\includegraphics[width=\textwidth]{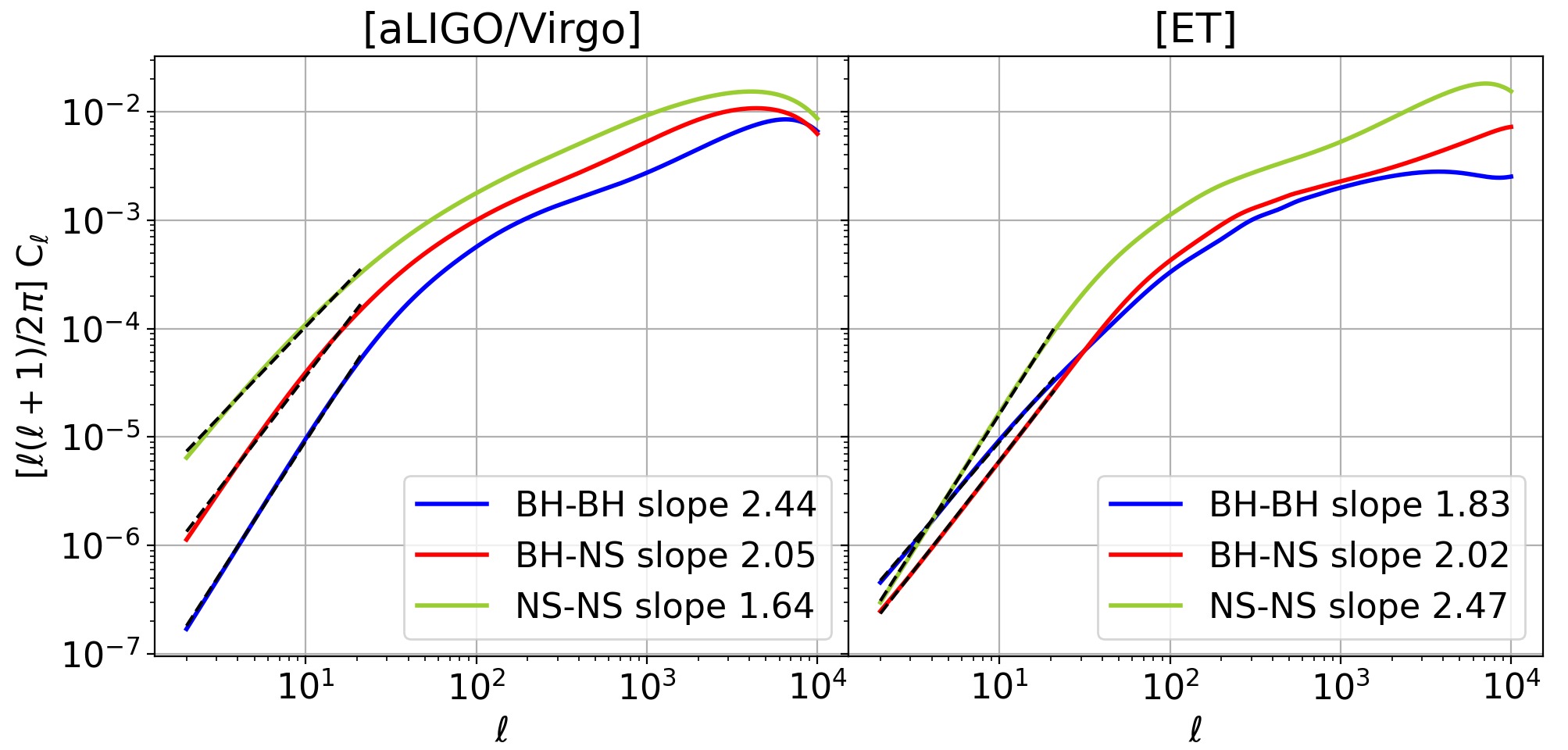}
\caption{Angular power spectrum of the anisotropies of the SGWB produced by BH-BH, BH-NS and NS-NS at 65 Hz as potentially measured by aLIGO/Virgo (left) and ET (right). The black dotted lines represent the power-law fit at large scales.}
\label{fig:cl_aligo_et}
\end{figure}

An important innovation of the present paper is the analysis of the statistical properties of the SGWB anisotropies as potentially measured by real detectors, i.e. considering only the unresolved GW events. The angular power spectrum of the SGWB anisotropies at 65 Hz for ET and aLIGO/Virgo are shown in figure \ref{fig:cl_aligo_et}.  
For both detectors and for all types of events, the power spectrum behaves like a power law at large angular scales that bends at smaller scales, as we also found when considering all the events. In this case, however, the the power law is steeper and its slope is close to 2. Actually, for both the considered detectors, the slope is not exactly equal to 2 but its value is slightly different depending on the type of source. We investigated the frequency behavior of the curves producing power spectra for $f = 33, 65, 105$ and $209$ Hz and we found that the slopes at large angular scales are constant with frequency. This is a valuable feature that could be used to distinguish the contribution of the various type of coalescing binaries from possible future measurements.
Interestingly, our framework produces predictions that are typical of each type of source and detector, as the outcome of the interplay of the different quantities involved in the calculation, namely the redshift distribution of the sources, their bias and magnification bias. In the following, we are going to give a qualitative interpretation of these features, with special focus on the large angular scales.

\begin{figure}
\centering
\subfloat[][\label{subfig:zcontr_all_bhbh}]
{\includegraphics[width=.49\textwidth]{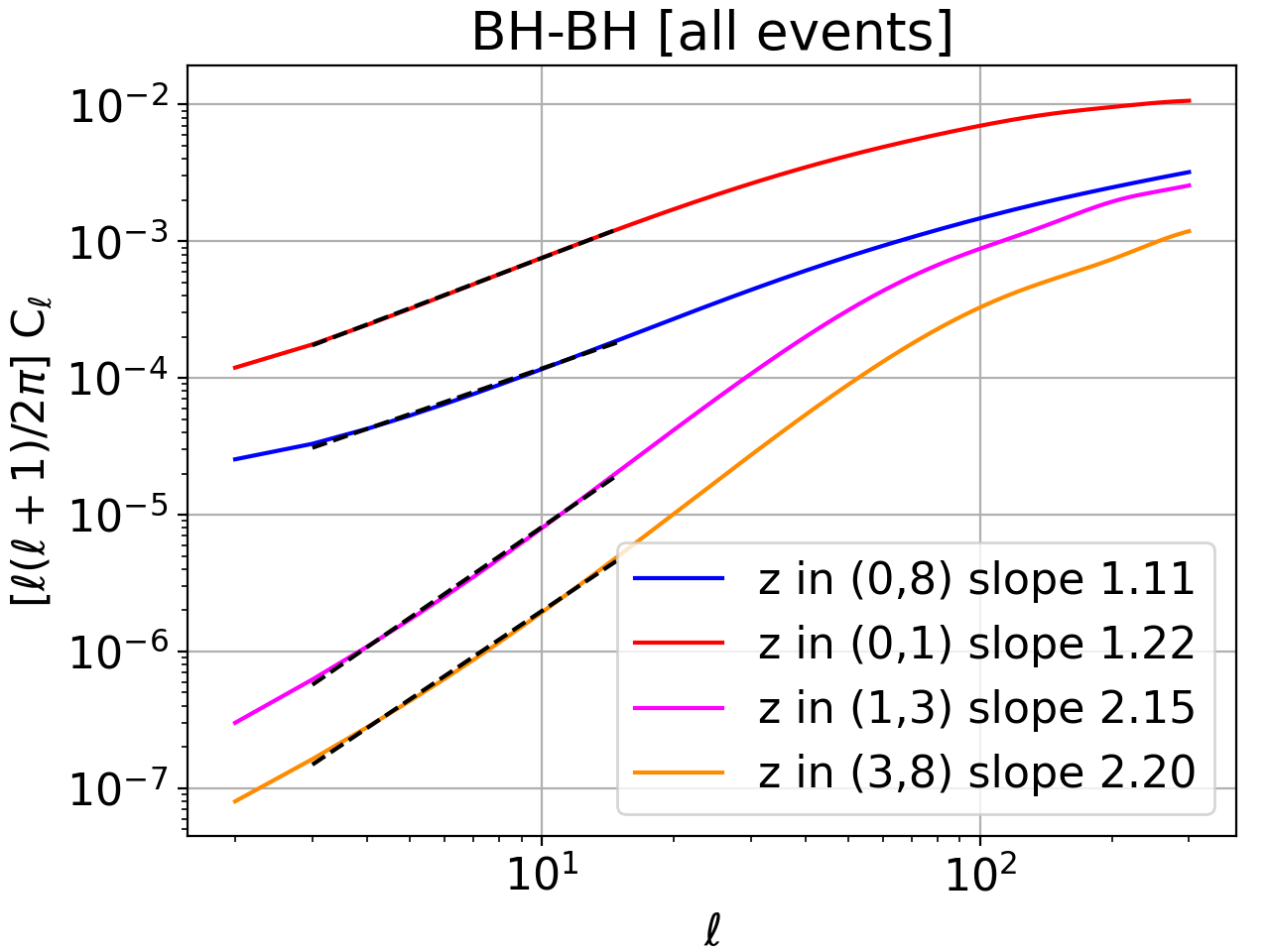}} \ 
\subfloat[][\label{subfig:zcontr_all_nsns}]
{\includegraphics[width=.49\textwidth]{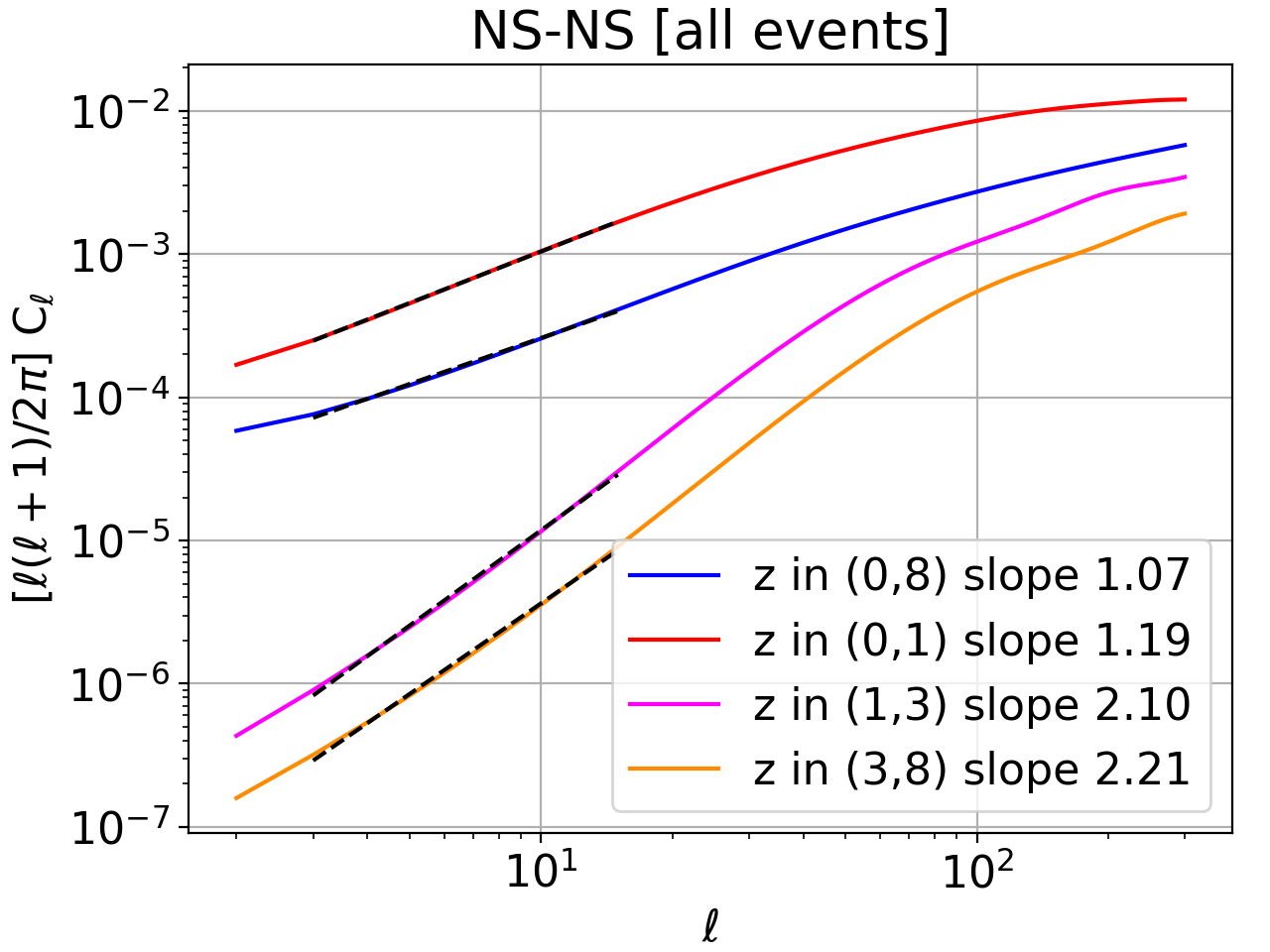}} \\
\subfloat[][\label{subfig:zcontr_aLIGO_bhbh}]
{\includegraphics[width=.49\textwidth]{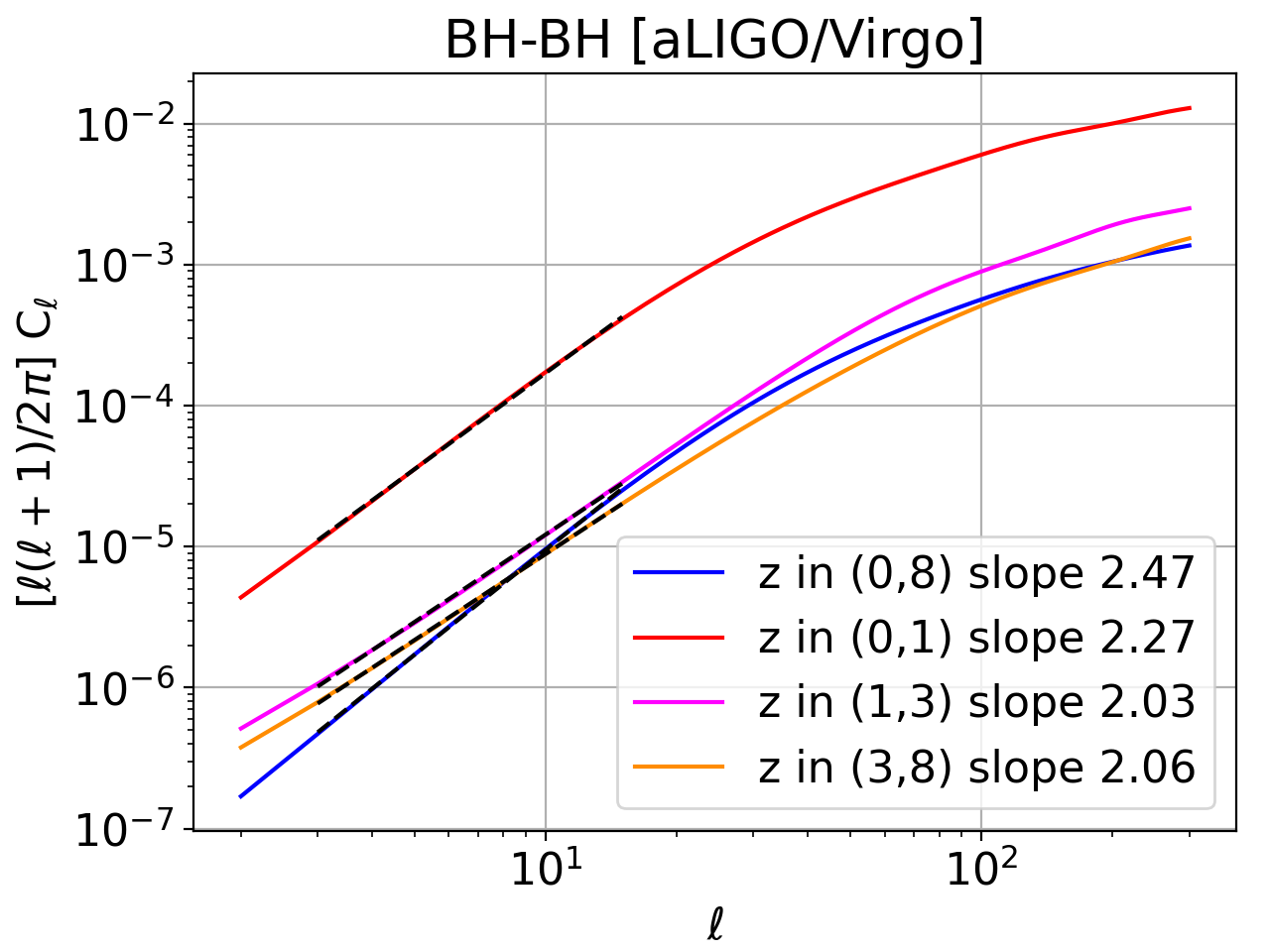}} \ 
\subfloat[][\label{subfig:zcontr_aLIGO_nsns}]
{\includegraphics[width=.49\textwidth]{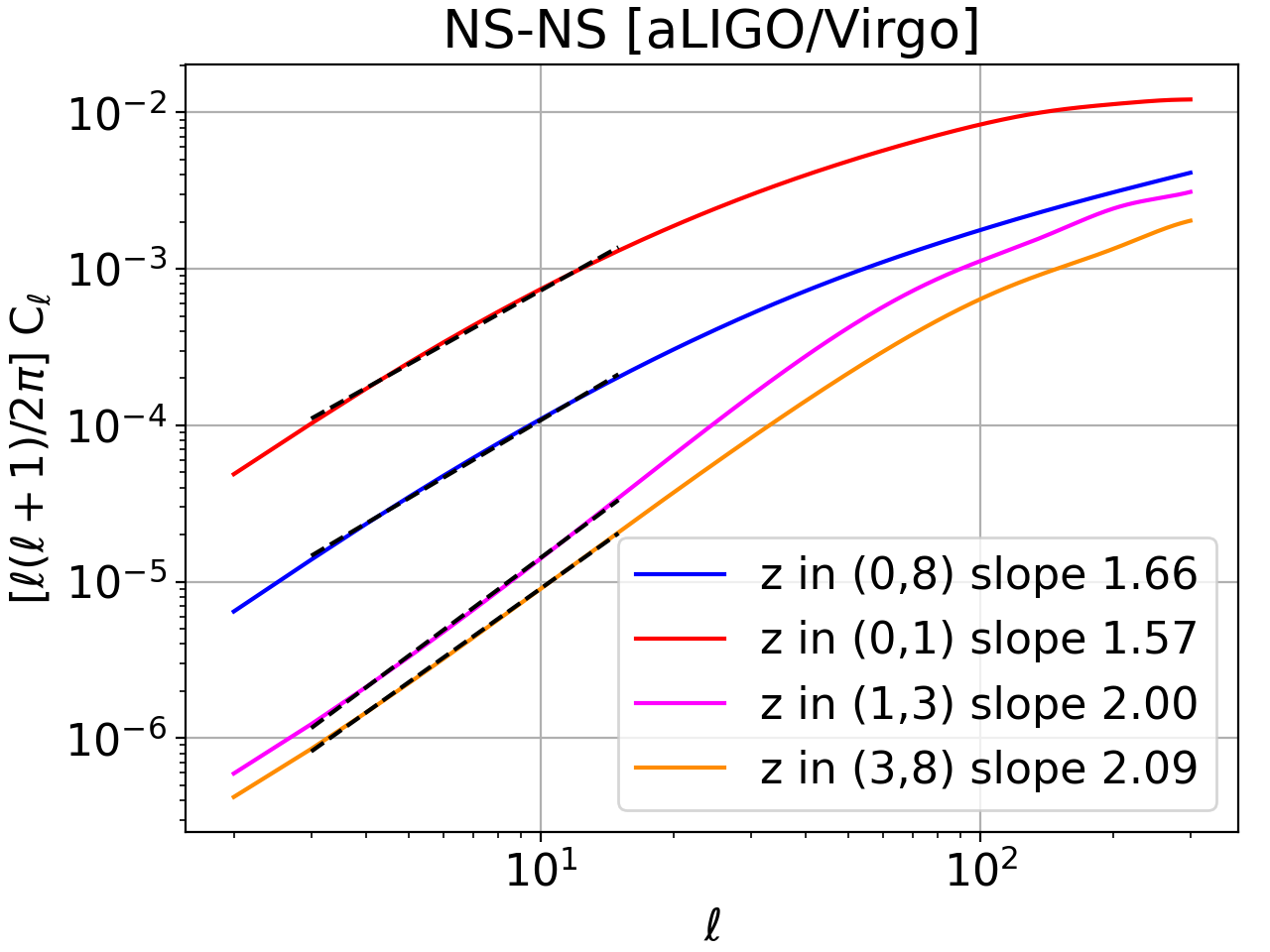}} \\
 \subfloat[][\label{subfig:zcontr_ET_bhbh}]
{\includegraphics[width=.49\textwidth]{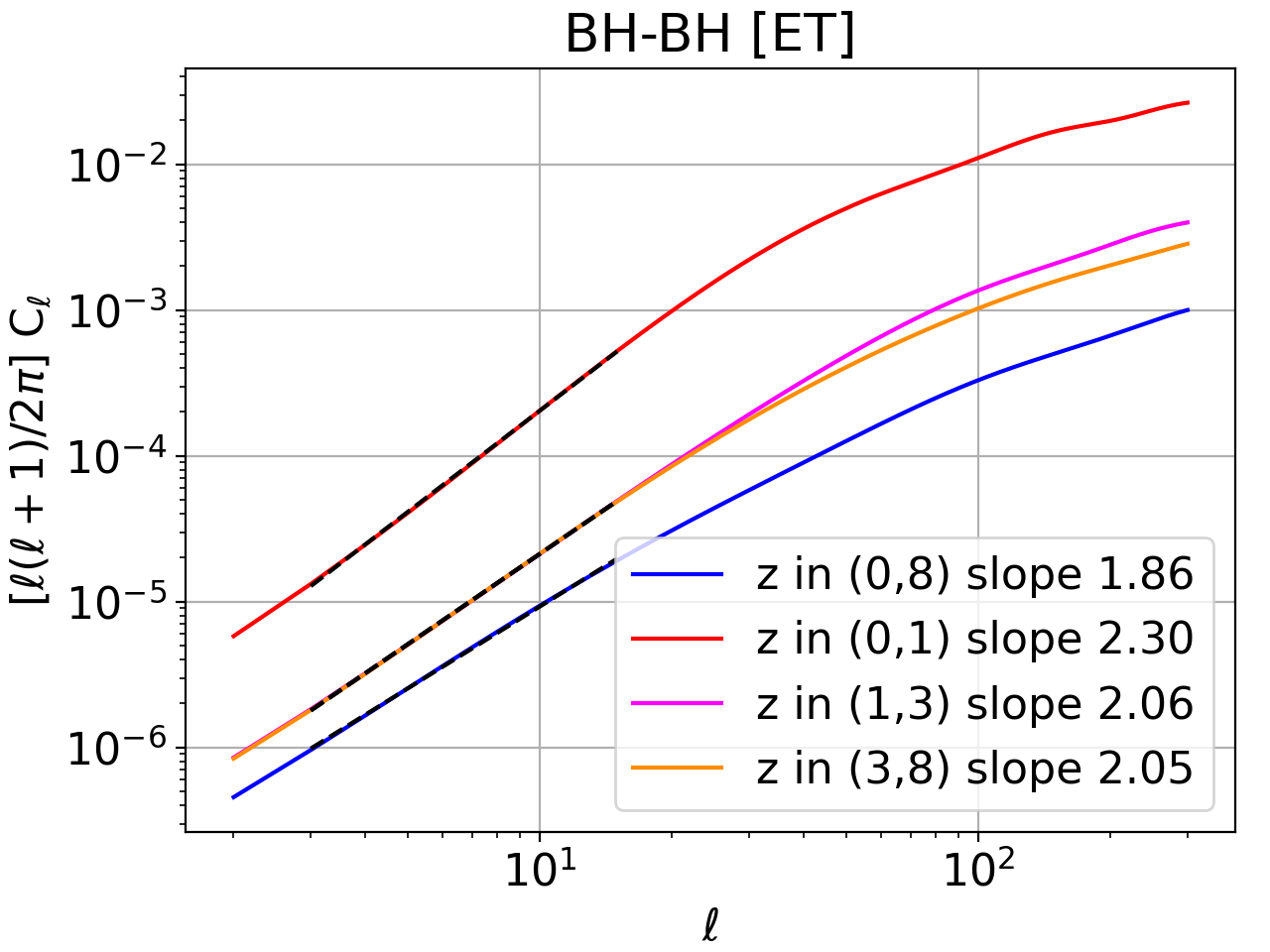}} \
\subfloat[][\label{subfig:zcontr_ET_nsns}]
{\includegraphics[width=.49\textwidth]{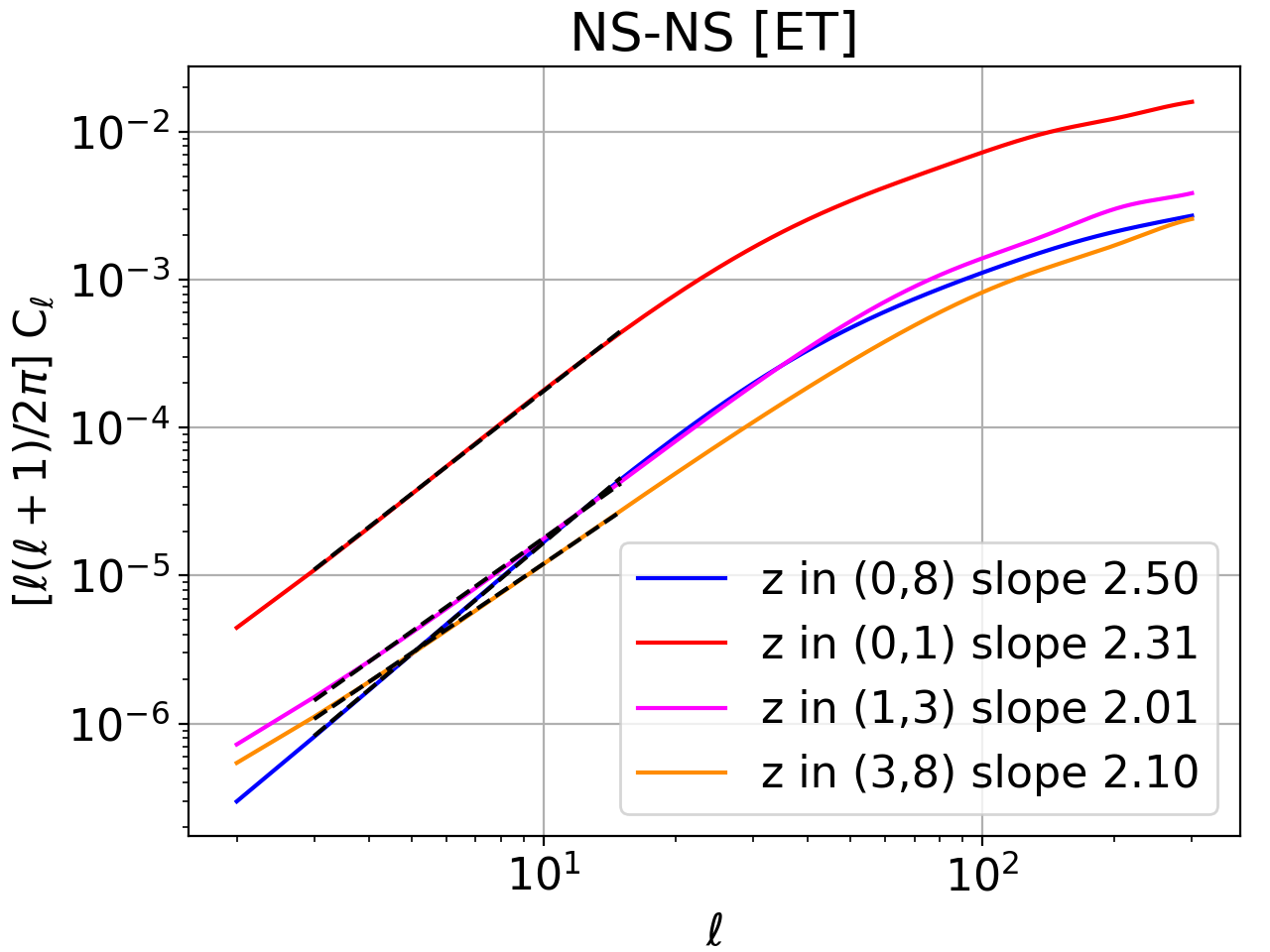}}
\caption{``Tomographic'' analysis applied to the anisotropies of the SGWB produced by BH-BH and NS-NS coalescing binaries at 65 Hz, considering all events (upper panels), for aLIGO/Virgo (middle panels) and for ET (lower panels). As explained in the text, the slope of the power spectrum at large angular scales allows to isolate the contributions of near and distant sources.}
\label{fig:z_contributions}
\end{figure}

The main difference when considering the anisotropies observed by real detectors is therefore a steeper power spectrum at large angular scales. The interpretation of this feature is not trivial: in fact, the SGWB is given by the superposition of all the unresolved GW events up to a typical maximum redshift $z_{\rm{max}}$, which is the maximum redshift at which sources can be observed by a given detector at a given frequency. It follows that there is no direct relation between wave modes and angular scales. Accordingly, a given wave-mode contributes to all multipoles such that $\ell \lesssim \ell_{\rm{max}}(k)$, where $\ell_{\rm{max}}(k)$ is related to $z_{\rm{max}}$ by
\begin{equation}
  \ell_{\rm{max}}(k) = k [\eta_{0} - \eta(z_{\rm{max}})]\, , 
\end{equation}
where $\eta(z)$ is the conformal time and $\eta_0 = \eta(0)$.
It is nevertheless possible to identify the contributions from events coming from different redshift ranges performing a sort of ``tomographic'' analysis that has to be interpreted in the light of the redshift distribution of the energy density $d\baromega/dz$. 
The analysis consists in computing the auto-correlation of the SGWB anisotropies in the redshift bins $z \in [0,1], [1,3]$ and $[3,8]$ and to compare them with the results in the total redshift interval $z \in [0,8]$. In figure \ref{fig:z_contributions}, we show the power spectra of the SGWB anisotropies in the selected redshift bins for BH-BH and NS-NS in the three different scenarios considered so far, i.e. SGWB produced by all events or only by unresolved events, for both aLIGO/Virgo and ET. 
We focus on the shape of the curves and do not consider their amplitude, which is typically higher for thinner redshift bins due to the normalization of the window functions and is not relevant for our current purposes.

Panels \ref{subfig:zcontr_all_bhbh} and  \ref{subfig:zcontr_all_nsns} show the results of the tomographic analysis for the SGWB produced by all the events. The behaviour of the curves is qualitatively the same for both BH-BH and NS-NS: the slope of the power-law at large angular scales is $\gtrsim 1$ for $z \in [0,1]$, whereas it is $\gtrsim 2$ for $z \in [1,3]$ and $z \in [3,8]$. We therefore deduce that the slope of the power-law is related to the distance of the considered sources. In particular, it is milder when the dominant contribution comes from low redshifts sources and steeper when the dominant contribution comes from high redshift sources. This is consistent with the fact that nearby events contribute to the power at large angular scales. Accordingly, since in this case the redshift distribution $d\omegagw/ dz$ is quite flat up to $z \sim 1$ for both BH-BH and NS-NS (see black curves in figure \ref{fig:domega_dz}) and nearby events contribute considerably to the signal, the slope of the power spectrum in the total redshift bin $z \in [0,8]$ is mild ($\gtrsim 1$) independently on the type of source (see also figure \ref{fig:Cls_all_events}).

Panels \ref{subfig:zcontr_aLIGO_bhbh} and  \ref{subfig:zcontr_aLIGO_nsns} show the power spectra of the anisotropies of the SGWB seen by aLIGO/Virgo. The situation is quite different in this case, since the detector will resolve many of the nearby events, so that the contribution of low redshift sources to the signal is suppressed, especially for BH-BH events (see blue curves in figure \ref{fig:domega_dz}). Consequently, the behavior at large scales is different for the two types of sources: in the NS-NS case, the power spectra are steeper (slope $\gtrsim 2$) for $z \in [1,3]$ and $z \in [3,8]$ and milder (slope $\sim 1.5$) for $z \in [0,1]$. In the BH-BH case, on the contrary, the slope is $\gtrsim 2$ in every redshift bin, because even for $z \in [0,1]$ the main contribution comes from events at $z \gtrsim 0.5$, since many nearby BH-BH events are resolved. First of all, these considerations explain why the power spectra of the SGWB anisotropies seen by aLIGO/Virgo in the total redshift interval $z \in [0,8]$ are generally steeper at large angular scales with respect to the case of the SGWB produced by all the events. Secondly, we deduce that the NS-NS component has a sightly milder slope than BH-NS and BH-BH (see figure \ref{fig:cl_aligo_et}) because it is made of a larger number low redshift events that give a higher contribution at large angular scales. As we already mentioned, the peculiar behavior of the different curves is the outcome of the combined intervention of many factors. However, for aLIGO/Virgo, the SGWB energy density is dominated by a population of events at relatively low redshift, so that the effects related to large scale structure and lensing are contained and the qualitative interpretation of the different slopes for different sources is reasonable.

Panels \ref{subfig:zcontr_ET_bhbh} and \ref{subfig:zcontr_ET_nsns}, finally, show the results for ET. In this case, the slope at large angular scales is $\sim 2$ in every redshift bin, independently of the type of source. The reason for this behavior is that the higher sensitivity of ET will allow to resolve almost all the nearby events, both BH-BH and NS-NS (see orange curves in \ref{fig:domega_dz}). This explains why the slopes of the power spectra are $\sim 2$ also for ET, but it is not sufficient to give an interpretation of the slightly different slopes that we obtain for different types of source (see figure \ref{fig:cl_aligo_et}). With respect to aLIGO/Virgo, in fact, ET is sensitive to a SGWB produced by a population of events at a longer distances, so that it is more difficult to keep track of all the different effects, related to large scale structure and lensing, that enter into play in the computation of the power spectrum.

We conclude with some considerations regarding the detectability of the anisotropic signal. First of all, we compare our results for with the current aLIGO/Virgo upper limit obtained during the first half of O3. In reference \citep{Abbott:2021jel} the results are given in terms of $\tilde{C}_{l}^{1/2} = \sqrt{C_{\ell}} \times \baromega / 4  \pi $ and the associated $95 \%$ upper limit at 25 Hz is $\tilde{C}_{\ell}^{1/2} \lesssim 1.9 \times 10^{-9} \; \rm{sr}^{-1}$ for $1\leq \ell \leq 4$, assuming an isotropic background with spectral index 2/3. Given our results for $\baromega$ and $C_{\ell}$ at 65 Hz for aLIGO/Virgo, we get $\tilde{C}_{\ell}^{1/2} \simeq 3.5 \times 10^{-14}\; \rm{sr}^{-1}$ at $\ell = 2$ for both BH-BH and NS-NS, which is far below the current upper limit.  
There are only a few forecasts for ET sensitivity to anisotropic SGWB. In reference \citep{Mentasti:2020yyd}, for example, the amplitude that the $\ell = 2$ multipole must have to produce a S/N = 1 in one year of observation of two ET-like detectors is found to be $\tilde{C}_{\ell}^{1/2} \lesssim 1.5 \times 10^{-14} \; \rm{sr}^{-1}$ at 10 Hz, assuming an isotropic background with spectral index 0. Our results for ET at 65 Hz lead to $\tilde{C}_{\ell}^{1/2} \simeq 6-8 \times 10^{-15}\; \rm{sr}^{-1}$ for the $\ell = 2$ multipole, depending on the type of merger, slightly below the minimum value in \citep{Mentasti:2020yyd}.

\subsection{SGWB anisotropy sky map}
\label{sec:sgwbasm}

\begin{figure}
    \centering
    \includegraphics[width=\textwidth]{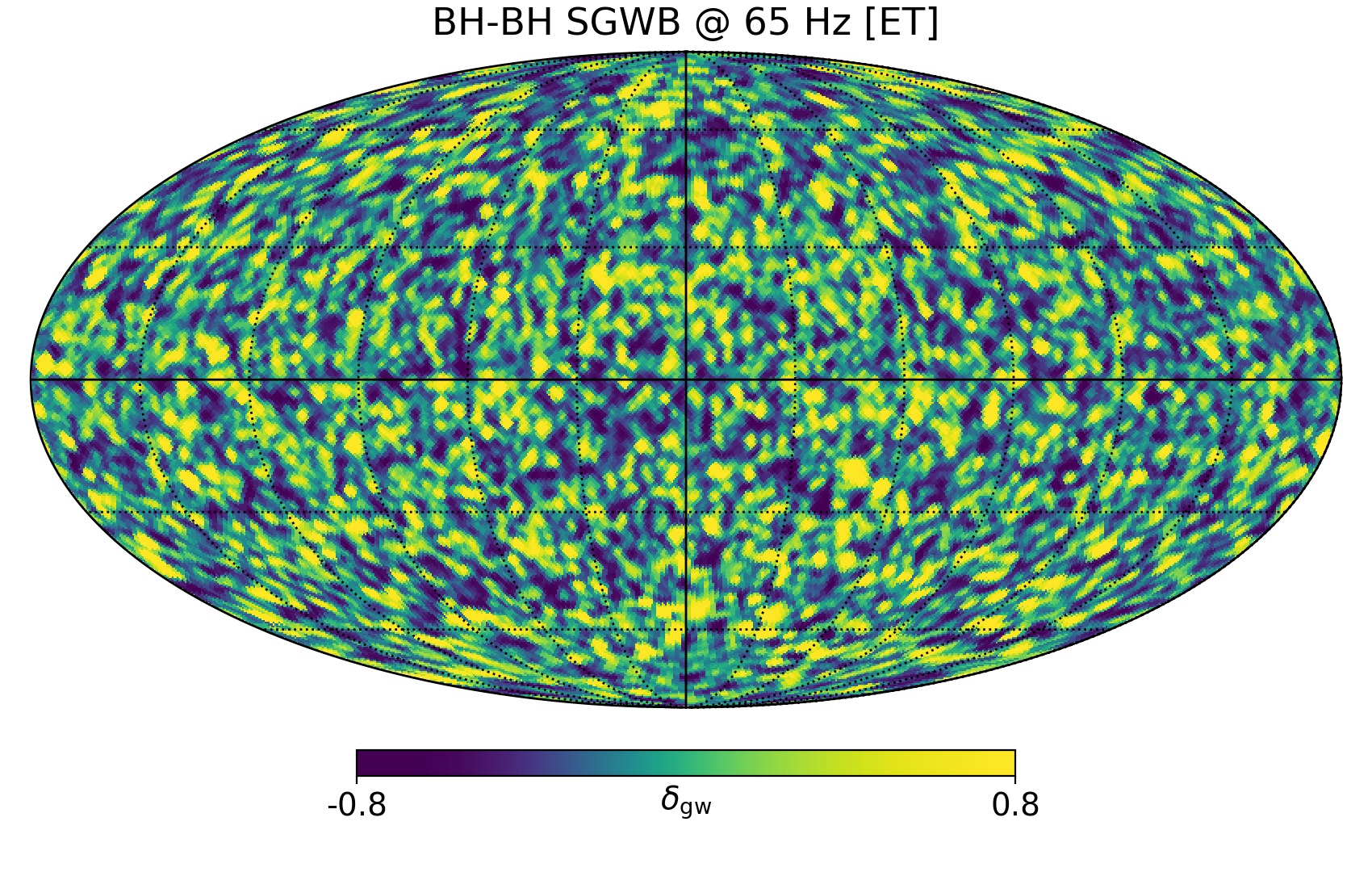}
    \includegraphics[width=\textwidth]{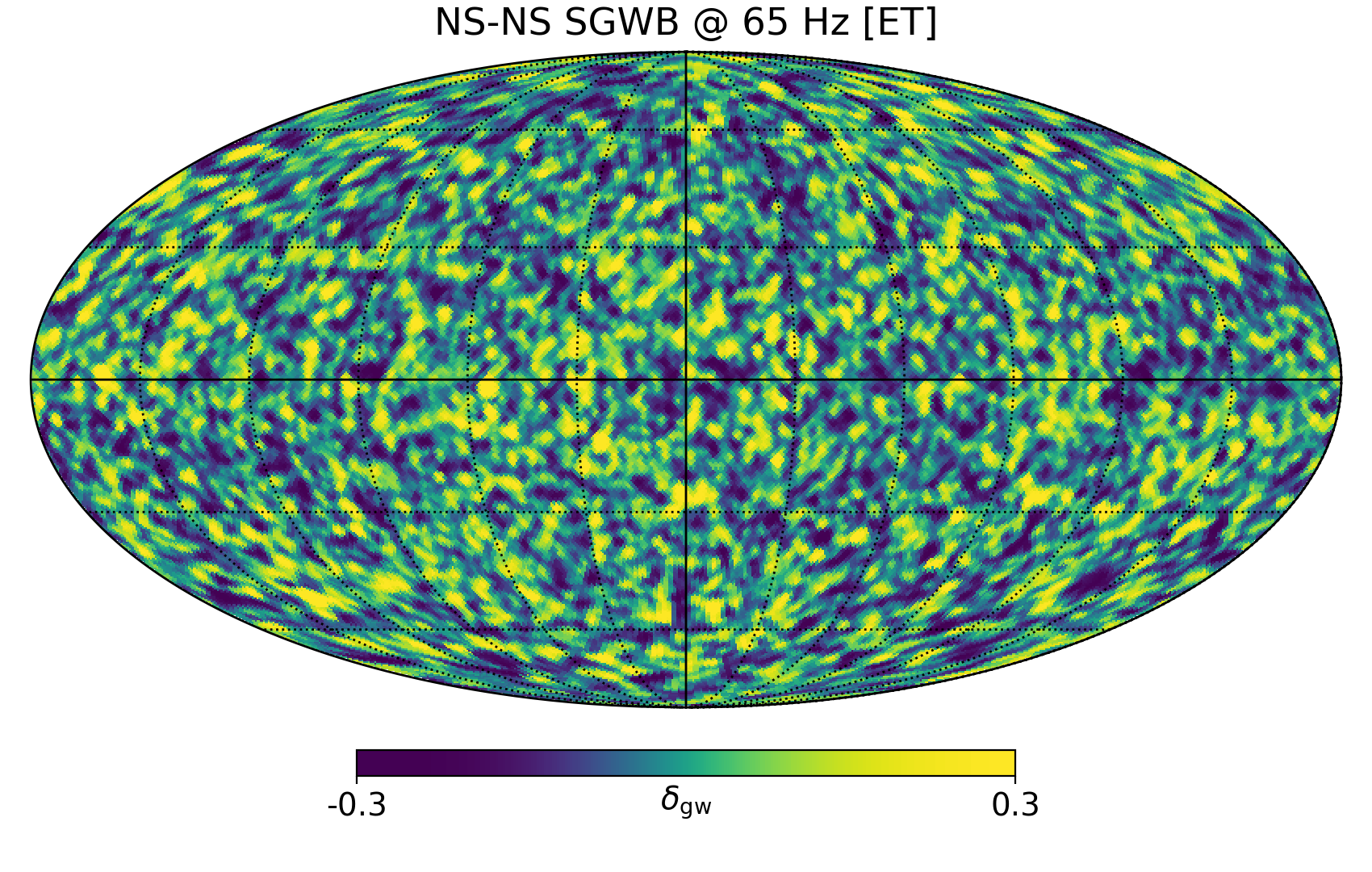}
    \caption{Mollweide maps of the energy density contrast of the SGWB produced by BH-BH (upper panel) and NS-NS (lower panel) mergers at 65 Hz for Einstein Telescope with integration time T = 1 yr. The plots have been realized with the HEALPix package, using $N_{\rm{side}} = 64$, which corresponds to an angular resolution of $\theta_{\rm{pix}} = 55$'. In order to remove the ringing effect arising when manipulating the harmonic coefficients, the maps have been smoothed with a Gaussian filter corresponding to $\sigma = \theta_{\rm{pix}}$.}
    \label{fig:maps}
\end{figure}

In this Sub-section, we exploit a map-making methodology in order to complement the power spectrum analysis above with realizations of the full sky anisotropy maps corresponding to the SGWB overdensity $\delta_{\rm{gw}}$. The peculiarity of the astrophysical SGWB with respect to other backgrounds, such as the Cosmic Microwave Background (CMB), is that it originates from unresolved point sources and hence its statistics follows a Poisson distribution instead of a Gaussian one.
We therefore need to produce a map of the anisotropies following a Poisson statistic with the cosmological clustering expected for the emitters that we obtained in the previous Sub-section. For this purpose, we adapt the procedure depicted in \citep{Gonzalez_Nuevo_2005} to the case of the SGWB. 

\begin{figure}
\includegraphics[width=\textwidth]{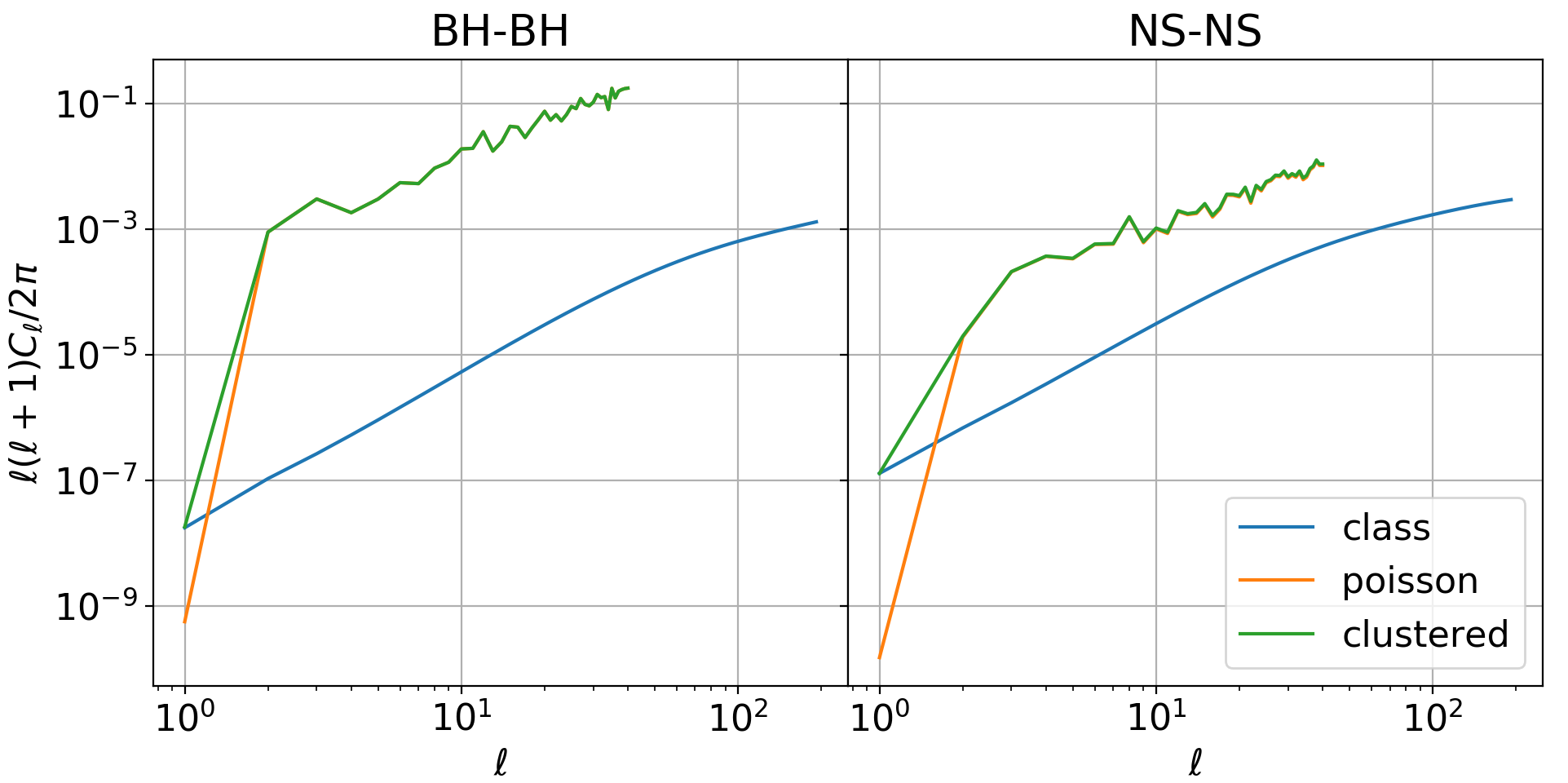}
\caption{Comparison among the power spectrum of the simulated SGWB maps (green), the theoretical power spectrum of the SGWB anisotropies obtained with \texttt{CLASS} (blue) and the power spectrum of pure Poisson noise (orange). The contribution of shot noise is $\propto \ell^{2}$ and it dominates over the signal at most scales. As expected, the noise is lower in the NS-NS case, since the merger rate is higher than the BH-BH one.}
\label{fig:cl_map}
\end{figure}

A preliminary step is to produce a Poisson map of the energy overdensity without any clustering. First of all, we compute the mean number of unresolved events per unit time \footnote{In our case, we count the events occurring in one year: this is tantamount to integrating the signal over $T=$ 1 yr. The choice of the integration time determines the mean number of events per pixel and hence the amplitude of the poissonian noise, as we will see in the following. See \citep{Jenkins:2019nks, Jenkins:2019uzp, Alonso:2020mva} for a detailed treatment of shot noise.} in each pixel:
\begin{equation}
    \langle \dot{N}_{\rm{pix}} \rangle = \dfrac{4 \pi}{N_{\rm{pix}}} \int dz \int d \mc \dfrac{d^{3} \dot{N}}{d \omega \, dz \, d\mc} \int_{0}^{\bar{\rho}} d \rho \, P_{\rho}(\rho| z, \mc)\,. 
\end{equation}
In the previous expression, $N_{\rm{pix}}$ is the number of pixels in the map and $d \dot{N}/d \omega \, dz \, d\mc$ is the merger rate per unit redshift, solid angle and chirp mass, which can be easily obtained from the differential merger rate of equation \eqref{eq:merger_rate} through: 
\begin{equation}
     \dfrac{d^{3} \dot{N}}{d \omega dz d\mc} = \dfrac{d^{2} \dot{N}}{dV d\mc}  \dfrac{c \, r(z)}{H_{0} \, h(z)}\,,
\end{equation}
where $r(z)$ is the comoving distance. We then create a map assigning to each pixel a number of events per unit time extracted from a Poisson distribution with mean $\langle \dot{N}_{\rm{pix}} \rangle$. We assign to each event in each pixel a chirp mass and a redshift that we generate randomly from a 2D probability distribution obtained from the differential merger rate $d \dot{N}/ dz \, d \mc$. In this way, we can compute the energy density in each pixel summing the contributions from all the events:
\begin{equation}
    \Omega^{\rm{Poiss}}_{\rm{gw}} = \dfrac{8 \pi G \fobs}{3 H_{0}^{2} c^{3}} \, \dfrac{1}{T} \, \sum_{i} \, \dfrac{\frac{dE}{df} (z_{i}, \mathcal{M}_{c \, i})}{4 \pi (1+z_{i}) r^{2}(z)}\,,
\end{equation}
where T is the considered unit of time (T = 1 yr in our case). At this stage, we have a map of the SGWB energy density which follows a pure Poisson statistics, without any clustering. In order to take the latter into account, first of all we compute the Poisson map of the density contrast:
\begin{equation}
    \delta^{\rm{Poiss}}_{\rm{gw}} = \dfrac{\omegagw^{\rm{Poiss}} - \langle \omegagw \rangle}{ \langle \omegagw \rangle}\,.
\end{equation}
Using the HEALPix\footnote{http://healpix.sourceforge.net} package \citep{Zonca2019,2005ApJ...622..759G}, we compute its harmonic coefficients $a_{\ell m}^{\rm{Poiss}}$. Subsequently, we introduce the correlation given by the angular power spectrum $C_{\ell}^{\rm{cl}}$ obtained with \texttt{CLASS} in the following way: 
\begin{equation}
    a_{\ell m}^{\rm{cl}} = a_{\ell m}^{\rm{Poiss}} \dfrac{\sqrt{C_{\ell}^{\rm{Poiss}} + C_{\ell}^{\rm{cl}}}}{\sqrt{C_{\ell}^{\rm{Poiss}}}}\,.
\end{equation}
Finally, by performing an inverse harmonic transform, we obtain the clustered Poisson density contrast $\delta_{\rm{gw}}^{\rm{cl}}$, from which the energy density at each pixel can be easily computed as $\omegagw^{\rm{cl}} = \langle \omegagw \rangle (1+ \delta_{\rm{gw}}^{\rm{cl}})$. 
In Figure \ref{fig:maps}, we show a full sky realization of the density contrast of the GW backgrounds produced by BH-BH and NS-NS mergers at 65 Hz for ET. In Figure \ref{fig:cl_map}, we plot the power spectra of the two maps, together with the theoretical power spectra obtained with \texttt{CLASS} and the power spectra of the pure poissonian maps (no clustering). As expected, the Poisson noise power spectrum is proportional to $\ell^{2}$ and its amplitude is related to the chosen integration time (one year in our case). By increasing the integration time, the number $N$ of events taken into account in each pixel would increase and the noise amplitude would decrease as $1/\sqrt{N}$. We find that for T = 1 yr the Poisson noise dominates over the signal, especially for the BH-BH component of the SGWB. The noise is lower for the NS-NS component because the merger rate - and hence the number of events per year in each pixel - is higher for NS-NS binaries than for BH-BH ones. In both maps the intrinsic correlation of the signal is higher than the Poisson noise at very large scales ($\ell = 1$), a feature that can also be seen observing the clustering in maps.   
From these results it is clear that being able to remove the shot noise component is of primary importance in order to extract a measurement of the power spectrum of the SGWB from real data. Recent studies address specifically this issue \citep{Jenkins:2019nks, Jenkins:2019uzp, Alonso:2020mva, Mukherjee:2019oma}: the effect of Poisson noise can be reduced increasing the integration time or cross-correlating the signal with other probes \citep{Cusin:2017fwz,Cusin:2019jpv,Canas-Herrera:2019npr}. A detailed analysis of the shot noise goes beyond the purposes of the present paper, but it is a crucial issue that will be surely addressed more deeply in future works.

\section{Conclusions}
\label{sec:conclusions}

In this paper, we studied the SGWB originated from the superposition of unresolved signals from merging compact binaries in galaxies, which is expected to be one of the dominant contributions of the total GW background in the LIGO/Virgo and ET frequency bands. The large amount of astrophysical and cosmological information encoded in this particular component of the SGWB makes its characterization one of the main goals for the GW community.
Our predictions rely on an empirical approach to galactic astrophysics that allows to follow the evolution of individual systems and on the results of the population synthesis code STARTRACK to model stellar and binary evolution.  

For the first time we have characterized the adimensional energy density parameter of the SGWB as a tracer of matter density, computing its redshift distribution, bias, magnification bias and evolution bias. With these elements and exploiting the similarities with the number counts formalism, we adapted the public code \texttt{CLASS} to the computation of the angular power spectrum of the SGWB anisotropies, taking into account lensing and relativistic corrections. We have obtained predictions concerning the isotropic energy density and the angular power spectrum of SGWB anisotropies for aLIGO/Virgo and ET. Furthermore, we also considered the case where all the events are taken into account, in order to compare with previous works in literature. All the results are diversified with respect to the various type of events (BH-BH, BH-NS and NS-NS) and the redshift location of the sources. 

For what concerns the isotropic component of the SGWB, we examined in detail the spectral shape of the energy density for all types of sources, comparing the results for the two detectors. In all cases, we recovered the well known power law behavior with spectral index $2/3$ at low frequencies with an exponential decrease at higher frequencies, depending on the typical chirp mass of the sources. In the case of the aLIGO/Virgo the dominating contribution to the energy density comes from BH-BH events, followed in order by BH-NS and NS-NS. For the ET, on the contrary, the contribution of NS-NS exceeds the one from BH-BH. This is due to the fact that ET is expected to resolve most of the low redshift BH-BH events. The better sensitivity of ET is also the reason why the overall energy of the SGWB is lower than in the aLIGO/Virgo case.

Regarding the SGWB anisotropies power spectrum, we found a power law behavior on large angular scales and a progressive drop at small scales, due to the lack of sources at very high redshifts. We performed a ``tomographic'' analysis in order to isolate the contribution of sources at different redshifts. We found that the slope of the power-law is related to nearby sources: the less they contribute, the steeper is the power spectrum. Accordingly, the anisotropies of the SGWB produced by the superposition of all the GW events, resolved and unresolved, have a power spectrum with a slope $\sim 1$ at large angular scales, whereas we found a slope $\sim 2$ when we used aLIGO/Virgo and ET sensitivities to subtract the resolved events, which are typically the nearest ones. We also found that the slopes of the power spectra are slightly different for different types of merger and detectors, and remain constant over a wide range of frequencies. This feature could be very useful in order to distinguish the different components of the SGWB from a possible future detection of the anisotropic signal. 

Finally, we presented a procedure to simulate realizations of the SGWB starting from the theoretical predictions of the power spectrum obtained with \texttt{CLASS}, and combining the Poisson nature of the statistics with the cosmological clustering properties. With this procedure, we obtained a high-resolution full sky map of the SGWB overdensity. The Poisson noise dominates over the pure clustering signal at all the considered angular scales, especially for the SGWB generated by BH-BH events, whose merger rate is lower than NS-NS ones.

In future works, we are going to extend the results of the present paper in a twofold way. First, we will exploit our analysis and techniques to cross-correlate the SGWB with other cosmological and astrophysical probes. This will be extremely useful in order to reduce the effects of Poisson noise and of systematic uncertainties, as well as to extract richer and more robust physical information. 
In addition, we will investigate how different stellar and galactic prescriptions affect our results. In fact, a detailed assessment of the modeling uncertainties in the astrophysical SGWB will constitute a primary issue for the component separation of the GW background, especially in the challenging search of the primordial contribution. 

\acknowledgments

A.L. acknowledges support from the PRIN MIUR 2017
prot. 20173ML3WW 002, `Opening the ALMA window on the cosmic evolution of gas, stars and supermassive black
holes' and the EU H2020-MSCA-ITN-2019 Project 860744 `BiD4BEST: Big Data applications for Black hole Evolution STudies'. G.C., G.S. and C.B. are partially supported by the INDARK INFN grant. C.B. acknowledges support from the COSMOS and LiteBIRD networks of the Italian Space Agency (www.cosmosnet.it). The authors would like to thank Daniele Bertacca, Nicola Bellomo, Enrico Barausse, Giulia Cusin for useful discussions and the anonimous referee for thoughtful evaluation and helpful suggestions given to improve our manuscript. Some of the results in this paper have been derived using the healpy and HEALPix packages \citep{Zonca2019,2005ApJ...622..759G}.

\bibliographystyle{unsrt}  
\bibliography{refs_sgwb}

\end{document}